\documentclass[a4paper,10pt]{article}
\usepackage[margin=1.2in]{geometry}

\usepackage{cancel,amsthm,amssymb,amsmath,mathrsfs,upgreek,stmaryrd,mathdots,graphicx,framed,verbatim,caption,subcaption,textcomp,color,tikz,todonotes}
\usetikzlibrary{patterns, decorations.markings,arrows, calc}
\usepackage[colorlinks=true, pdfstartview=FitV, linkcolor=darkblue, citecolor=darkblue, urlcolor=darkblue]{hyperref}

\definecolor{shadecolor}{rgb}{0.9, 0.9, 0.86}
\definecolor{darkblue}{rgb}{0.1,0.1,0.45}

\def\Re{\mathrm {Re}\,}
\def\Im{\mathrm {Im}\,}
\def \wt{\widetilde}
\def \wh{\widehat}
\newcommand{\C}{\mathbb{C}}

\newcommand{\R}{\mathbb{R}}
\newcommand{\Z}{\mathcal{Z}}
\newcommand{\G}{\Gamma}
\newcommand{\1}{\mathbf{1}}
\newcommand{\J}{\mathcal{J}}
\renewcommand{\H}{\mathcal{H}}

\newcommand{\E}{\mathrm{E}}

\renewcommand{\a}{\alpha}

\renewcommand{\L}{\Lambda}
\renewcommand{\d}{\mathrm d}
\renewcommand{\O}{\mathcal{O}}
\renewcommand{\S}{\Sigma}
\def \SS {\mathcal{S}}
\newcommand{\U}{\mathrm{U}}

\newcommand{\e}{\mathrm{e}}
\renewcommand{\i}{\mathrm{i}}

\newcommand{\pa}{\partial}
\newcommand{\z}{\mathbf{z}}
\def\res{\mathop{\mathrm {res}}\limits_}
\def \tr {\mathrm{tr}\,}
\def\be{\begin{equation}}
\def\ee{\end{equation}}
\def\bg{\begin{gathered}}
\def\eg{\end{gathered}}
\def\Ai{\mathrm{Ai}}
\def\K{\mathrm{K}}
\def \U{\mathrm{U}}

\newtheorem{theorem}{Theorem}[section]

\newtheorem{lemma}[theorem]{Lemma}
\newtheorem{remark}[theorem]{Remark}
\newtheorem{problem}[theorem]{Riemann-Hilbert Problem}

\newtheorem{proposition}[theorem]{Proposition}

\def\QED {\hfill $\blacksquare$\par\vskip 3pt}
\def\s{\sigma}
\def\bea{\begin{eqnarray}}
\def\eea{\end{eqnarray}}
\def\Mgn{\overline{\mathcal{M}}_{g,n}}
\def\P{\mathbb{P}}
\def\ev{\mathrm{ev}}

\def\T{\mathbf{T}}
\def\J{\mathrm{J}}
\def\Y{\mathrm{Y}}
\def\H{\mathrm{H}}
\def\ss{\mathbf{s}}

\DeclareMathOperator{\SL}{SL}

\DeclareMathOperator{\diag}{diag}

\renewcommand{\theequation}{\arabic{section}.\arabic{equation}}

\makeatletter
\@addtoreset{equation}{section}
\makeatother

\begin{document}

\vspace{0.2cm}
\begin{center}
\begin{Large}
\textbf{Matrix models for stationary Gromov-Witten invariants of the Riemann sphere} 
\end{Large}
\end{center}

\bigskip
\begin{center}
M. Bertola$^{\dagger\ddagger\clubsuit}$\footnote{Marco.Bertola@\{concordia.ca, sissa.it\}},  
G. Ruzza $^{\diamondsuit}$ \footnote{giulio.ruzza@uclouvain.be}.
\\
\bigskip
\begin{minipage}{0.7\textwidth}
\begin{small}
\begin{enumerate}
\item [${\dagger}$] {\it  Department of Mathematics and
Statistics, Concordia University\\ 1455 de Maisonneuve W., Montr\'eal, Qu\'ebec,
Canada H3G 1M8} 
\item[${\ddagger}$] {\it SISSA, International School for Advanced Studies, via Bonomea 265, Trieste, Italy }
\item[${\clubsuit}$] {\it Centre de recherches math\'ematiques,
Universit\'e de Montr\'eal\\ C.~P.~6128, succ. centre ville, Montr\'eal,
Qu\'ebec, Canada H3C 3J7} 
\item[${\diamondsuit}$] {\it Institut de recherche en math\'{e}matique et physique,
Universit\'e catholique de Louvain, Chemin du Cyclotron 2,
1348 Louvain-la-Neuve, Belgium} 
\end{enumerate}
\end{small}
\end{minipage}
\vspace{0.5cm}
\end{center}
\bigskip

\begin{center}
\begin{abstract}
Inspired by recent formul\ae\ of Dubrovin, Yang, and Zagier, we interpret the tau function enumerating stationary Gromov-Witten invariants of $\P^1$ as an isomonodromic tau function associated with a difference equation. As a byproduct we obtain an analogue of the Kontsevich matrix model for this tau function. A connection with the Charlier ensemble is also considered.
\end{abstract}

\end{center}

\setcounter{tocdepth}{1}

\tableofcontents

\section{Introduction and results}

The well known conjecture by  Witten \cite{Wi1991} and subsequent proof by Kontsevich \cite{Ko1992} says that if we consider the  following generating function of intersection numbers on $\Mgn$ 
\be
\label{F}
F(t_0,t_1,t_2,...):=\sum_{\begin{smallmatrix}g,n\geq 0 \\ 2g-2+n>0\end{smallmatrix}}\sum_{k_1,...,k_n\geq 0}
\frac{t_{k_1}\cdots t_{k_n}}{n!}
\int_{\Mgn}\psi_1^{k_1}\cdots\psi_n^{k_n}
=\frac{t_0^3}6+\frac{t_1}{24}+\frac{t_0^3t_1}{24}+\frac{t_{0}t_{2}}{24}+\frac{t_{1}^{2}}{24}+\cdots
\ee
then $\exp F$ is a tau function of the KdV hierarchy, termed \emph{Kontsevich-Witten tau function}.
For the reader unfamiliar with these notions from algebraic geometry we have included a small informal introduction explaining all the terminology in Appendix \ref{appb}.
In \eqref{F}, $\Mgn$ is the Deligne-Mumford compactification of the moduli space of Riemann surfaces with $n$ marked points and $\psi_i$ is the Chern class of the cotangent line bundle at the $i$th puncture, $i=1,...,n$. The dimensional constraint $k_1+...+k_n=3g-3+n$ allows to read off the corresponding  genus $g$ for every coefficient of the generating function \eqref{F}. 

The Kontsevich-Witten tau function is closely related to the \emph{Kontsevich matrix model}
\be
\label{zk}
\Z^{Kont}_N(z_1,...,z_N):=\frac{\int_{\H_N}\exp\tr(\i\frac{M^3}3-Z M^2)\d M}{\int_{\H_N}\exp\tr(-ZM^2)\d M}=\frac{\det\left(\sqrt{4\pi z_j}\e^{\frac 23 z_j^3}\Ai^{(k-1)}(z_j^2)\right)_{j,k=1}^N}{\Delta(-z_1,...,-z_N)}
\ee
where $\H_N$ is the space of $N\times N$ hermitian matrices, $Z=\diag(z_1,...,z_N)$ is an $N\times N$ diagonal matrix, and $\Ai(\zeta)=\frac 1 \pi \int_{0}^{+\infty}\cos\left(\frac{x^3}3+\zeta x\right)\d x$ is the standard Airy function. Here and elsewhere we denote 
\be
\Delta(x_1,...,x_m):=\det\left(x_j^{k-1}\right)_{j,k=1}^m=\prod_{1\leq j<k\leq m}(x_k-x_j)
\ee
the Vandermonde determinant. The proof of the equality in \eqref{zk} uses standard techniques of matrix integration, and is completely analogous to the arguments which we will use to prove \eqref{zvafa} below.
The connection between \eqref{F} and \eqref{zk} goes as follows. The Airy function admits an asymptotic expansion $\sqrt{4\pi z}\e^{\frac 23 z^3}\Ai(z^2)\sim\sum_{j\geq 0}\frac{(6j-1)!!}{(2j)!72^j}\frac{(-1)^j}{z^{3j}}$ for $z\to\infty$ within $|\arg z_j|<\frac\pi 2$, and so \eqref{zk} admits an asymptotic expansion of the form $1+\O(z^{-1}_j)$ which is a symmetric formal power series in $z_1^{-1},...,z_N^{-1}$. Consider this expansion for \eqref{zk} in terms of the scaled {\it Miwa times} $t_k:=-(2k-1)!!\,\tr\left[ \left(\sqrt[3]{2}Z\right)^{-2k-1}\right]$, which is a formal power series in $t_0,t_{\frac 12},t_1,...$. Setting $\deg t_k:=2k+1$, it can be shown that terms up to degree $D$ in this expansion for $\Z^{Kont}_N(z_1,...,z_N)$ do not depend on $N$ as soon as $N>D$, in other terms they \emph{stabilize} as $N\to\infty$; moreover, coefficients in front of monomials involving $t_k$'s with non-integer indexes vanish \cite{ItZu1992,Di2003}. Finally, the logarithm of this limiting expansion coincides with the generating function \eqref{F} \cite{Ko1992}.

Generalizations of this result in various directions have been considered; in particular to $r$-spin intersection numbers \cite{Wi1993,AdVM1992}, to open intersection numbers \cite{BrHi2012,AlBuTe2017,Aleopen1,Aleopen2}, and to Gromov-Witten (GW) theory \cite{Kontsevich1994,Behrend1996}. 

One of the first important examples of the last case is the stationary GW theory of $\P^1$. In this case, the generating function \eqref{F} is replaced by
\begin{align}
\nonumber
F_{\P^1}(t_0,t_1,t_2,...;\epsilon)&:=\sum_{g,n\geq 0}
\sum_{k_1,...,k_n\geq 0}
\frac{t_{k_1}\cdots t_{k_n}}{n!}\epsilon^{2g-2}
\int_{\left[\Mgn(\P^1;d)\right]}\psi_1^{k_1}\cdots\psi_n^{k_n}\ev_1^*\omega\cdots\ev_n^*\omega
\\
\label{FP1}
&=\left(\frac 1{\epsilon^2}-\frac 1{24}\right)t_0+\frac {t_0^2}{2\epsilon^2}+\frac{t_0^3}{6\epsilon^2}+\left(\frac 1{4\epsilon^2}+\frac 1{24}+\frac {7\epsilon^2}{5760}\right)t_2+\cdots
\end{align}
where $\Mgn(\P^1;d)$ denotes the moduli space of degree $d$ stable maps from Riemann surfaces of genus $g$ with $n$ marked points to $\P^1$; $\left[\Mgn(\P^1;d)\right]$ is the virtual fundamental class \cite{Behrend1997}, which allows integration of characteristic classes, in this case the psi-classes $\psi_i$ as above (pulled back via the forgetful map $\Mgn(\P^1;d)\to\Mgn$) and the classes $\ev_i^*\omega$ (pullback of the normalized K\"{a}hler class $\omega\in H^2(\P^1;\mathbb Z)$, $\int_{\P^1}\omega=1$, via the evaluation maps $\ev_i:\Mgn(\P^1;d)\to\P^1$ at the $i$th marked point).

The dimensional constraint $k_1+...+k_n=2(g-1+d)$ allows to recover the degree $d$ for every coefficient of the generating function \eqref{FP1}. The exponential $\exp F_{\P^1}$ is a tau function of the Toda hierarchy \cite{OkPa2006,DuZh2004}.

Introduce the following entire function $f(z;\epsilon)$ of the complex variable $z$, depending on a parameter $\epsilon$ which will be assumed real positive for the rest of this work, $\epsilon>0$;
\be
f(z;\epsilon):=\frac 1{\sqrt{2\pi\epsilon}}\int_{C_1}\exp\left(\frac 1\epsilon\left(x-\frac 1x\right)-\left(z+\frac 32\right)\log x\right) \d x.
\ee
The contour $C_1$ starts from $0$ with $|\arg x|<\frac\pi 2$ and arrives at $\infty$ with $\frac \pi 2<\arg x<\pi$, see Fig. \ref{figurecontour} below. The function $f$ is a Hankel function of the argument, see Rem. \ref{rembessel} and  Sec. \ref{secbare}. Define
\be
\label{zour}
\Z_N(z_1,...,z_N):=
\frac{\det\left(\frac 1{\epsilon^{k-1}}\left(\frac{\epsilon z_j}\e\right)^{-z_j}f(z_j+k-1;\epsilon)\right)_{j,k=1}^N}{\Delta(z_1,...,z_N)}.
\ee
We will explain the connection of the function   $\Z_N$  to a suitable matrix model below (see \eqref{matrixmodel}).

The function  $f(z;\epsilon)$ satisfies the second order difference equation
\be
\label{difference}
f(z+1;\epsilon)+f(z-1;\epsilon)=\epsilon\left(z+\frac 12 \right)f(z;\epsilon)
\ee
and admits an asymptotic expansion within the sector $|\arg z|\leq \frac \pi 2-\delta$, $\delta>0$, of the form
\be
\label{besselasy}
\left(\frac{\epsilon z}{\e}\right)^{-z}f(z;\epsilon)\sim 
1+\frac{24-\epsilon^2}{24 \epsilon^2z}+\frac{\epsilon^4+528 \epsilon^2+576}{1152 \epsilon^4z^2}+\frac{1003 \epsilon^6+95400 \epsilon^4+406080 \epsilon^2+69120}{414720 \epsilon^6z^3}+...
\ee
where  the coefficients can be computed either by a steepest descent analysis or by the difference equation \eqref{difference}; these statements are proven below in Sec. \ref{secbare}. Therefore, within the same sector we also have
\be
\left(\frac{\epsilon z}{\e}\right)^{-z}f(z+k;\epsilon)\sim (\epsilon z)^k (1+\O(z^{-1})).
\ee
for all $k=0,1,2,...$. This implies that the ratio \eqref{zour} admits, in the same sector, an asymptotic expansion of the form $1+\O(z^{-1}_j)$; this expansion for \eqref{zour} is a symmetric formal power series in $z_1^{-1},...,z_N^{-1}$, which stabilizes once expressed in terms of the scaled Miwa variables
\be
\label{scaledmiwa}
t_k:=\frac{k!}{\epsilon^k}\left(\frac 1{z_1^{k+1}}+\cdots+\frac 1{z_N^{k+1}}\right).
\ee
Namely, setting $\deg t_k:=k+1$, terms of degree $D$ in the expansion of $\Z_N(z_1,...,z_N)$ do not depend on $N$ as soon as $N>D$; the proof is exactly the same as for the Kontsevich model \cite{ItZu1992,Di2003}.

Our main result is that \eqref{zour} is the correct analogue of the Kontsevich model for the stationary GW theory of $\P^1$.

\begin{shaded}
\begin{theorem}
\label{mainthm}
The expansion of $\log \Z_N(z_1,...,z_N)$, expressed in terms of the scaled Miwa variables \eqref{scaledmiwa}, stabilizes as $N\to\infty$ to the generating function \eqref{FP1} of stationary GW invariants of $\P^1$. 
\end{theorem}
\end{shaded}

For example, using the first terms of the expansion in \eqref{besselasy} we can compute $\Z_{N=4}(z_1,...,z_4)$ up to terms of order $3$ in $z_1^{-1},...,z_4^{-1}$ as
\begin{align*}
\Z_{N=4}(z_1,...,z_4)&=\scriptstyle{1+\frac{24-\epsilon^2}{24\epsilon^2}\left(\frac 1{z_1}+\cdots+\frac 1{z_4}\right)+\left(\frac 1{1152}+\frac{11}{24\epsilon^2}+\frac 1{2\epsilon^4}\right)\left(\frac 1{z_1^2}+\cdots+\frac 1{z_4^2}+\frac 2{z_1z_2}+\cdots+\frac 2{z_3z_4}\right)}
\\
&
\!\!\!\!\!\!\!\!\!\!\!\!\!\!\!\!\!\!\!\!\!\!\!\!\!\!\!\!\!\!\!\!\!\!\!\!\!\!\!\!\!\!
\scriptstyle{+\left(\frac{1003}{414720}+\frac{265}{1152\epsilon^2}+\frac{47}{48\epsilon^4}+\frac 1{6\epsilon^6}\right)\left(\frac 1{z_1^3}+\cdots+\frac 1{z_4^3}\right)
+\left(-\frac 1{27648} + \frac{169}{384\epsilon^2}+\frac{23}{16\epsilon^4}+\frac 1{2\epsilon^6}\right)\left(\frac 1{z_1^2z_2}+\cdots+\frac 1{z_3z_4^2}+\frac 2{z_1z_2z_3}+\cdots+\frac 2{z_2z_3z_4}\right)
}
\end{align*}
and then, in view of the relations
\begin{align*}
t_0&=\frac 1{z_1}+\cdots+\frac 1{z_4},
&
t_0^2&=\frac 1{z_1^2}+\cdots+\frac 1{z_4^2}+\frac 2{z_1z_2}+\cdots+\frac 2{z_3z_4},
\\
\frac{\epsilon^2 t_2}2&=\frac 1{z_1^3}+\cdots+\frac 1{z_4^3},
&
\frac{t_0^3}3-\frac{\epsilon^2t_2}{6}&=\frac 1{z_1^2z_2}+\cdots+\frac 1{z_3z_4^2}+\frac 2{z_1z_2z_3}+\cdots+\frac 2{z_2z_3z_4},
\end{align*}
the expansion for $\log \Z_{N=4}(z_1,...,z_4)$ correctly reproduces the terms up to degree $3$ given by example in \eqref{FP1}.

\subsection{Connections with matrix models}

\paragraph*{Matrix models with external source.}

A $\cosh$-potential for a Kontsevich matrix model for stationary GW invariants of $\P^1$ has been proposed in \cite{Vafa}; in the naive interpretation with a flat hermitian measure this matrix model reads
\footnote{More precisely, in \cite{Vafa} the integrand of the matrix model partition function is identified as $\exp\frac 1g \tr\left(M\L-\e^M-q\e^{-M}\right)$. Up to minor modifications, the parameters $g,q$ can be combined into a single parameter $\epsilon=q^{-\frac 12}g$; then the integrand of \eqref{zvafa} is recovered by the identification $\L=g Z$.}
\footnote{However in \cite{Vafa} the measure considered is not identified as the flat measure $\d M$. We thank Prof. A. Alexandrov for pointing this out.}
\be
\label{zvafa}
\int_{\H_N}\exp\tr\left(MZ-\frac 2\epsilon\cosh M\right)\d M=\pi^{\frac{N(N-1)}2}\frac{\det\left(\int_\R x^k\e^{xz_j-\frac 2\epsilon\cosh x}\d x\right)_{j,k=1}^N}{\Delta(z_1,...,z_N)}.
\ee

The equality above can be derived as follows. First we decompose integration in eigenvalues and angular variables
\begin{align}
\int_{\H_N}\exp\tr\left(MZ-\frac 2\epsilon\cosh M\right)\d M=\frac {\pi^{\frac{N(N-1)}2}}{\prod_{i=1}^Ni!}\int_{\R^N}\left(\int_{\U_N}\e^{\tr\left(UXU^\dagger Z\right)}\d U\right)\Delta^2(X)\e^{-\frac 2\epsilon\tr\cosh(X)}\d X
\end{align}
denoting $\d U$ the normalized Haar measure over the unitary group $\U_N$ of $N\times N$ matrices, $\int_{\U_N}\d U=1$, $\d X=\d x_1...\d x_N$, and $\Delta(X):=\Delta(x_1,...,x_N)$. Then we use Harish--Chandra-Itzykson-Zuber formula
\be
\int_{\U_N}\e^{\tr\left(UXU^\dagger Z\right)}\d U=\left(\prod_{i=1}^{N-1}i!\right)\frac{\det\left(\e^{x_iz_j}\right)_{i,j=1}^N}{\Delta(X)\Delta(Z)}
\ee
to rewrite the previous expression as
\be
\frac{\pi^{\frac{N(N-1)}2}}{N!}\frac{1}{\Delta(Z)}\int_{\R^N}\Delta(X)\det\left(\e^{x_iz_j-\frac 2\epsilon\cosh x_i}\right)_{i,j=1}^N\d X
\ee
and finally the equality in \eqref{zvafa} is a consequence of the \emph{Andreief identity}
\be
\label{andreief}
\int_{\R^N}\det\left(\phi_i(x_j)\right)_{i,j=1}^N\det\left(\psi_i(x_j)\right)_{i,j=1}^N\d x_1....\d x_N=N!\det\left(\int_{\R}\phi_i(x)\psi_j(x)\d x\right)_{i,j=1}^N
\ee
with $\phi_i:=x^{i-1}$ and $\psi_i(x):=\e^{xz_i-\frac 2\epsilon\cosh x}$.
Noting now that
\be
\int_\R \e^{xz-\frac 2\epsilon\cosh x}\d x=2\K_{z}\left(\frac 2\epsilon\right)
\ee
where $\K_\nu(\zeta)$ is the modified Bessel function of second kind of order $\nu$ and argument $\zeta$ \cite{AbSt1965}, the matrix integral \eqref{zvafa} can be alternatively expressed as
\be
2^N\pi^{\frac{N(N-1)}2}\frac{\det\left(\pa^{k-1}_{z_j}\K_{z_j}\left(\frac 2\epsilon\right)\right)_{j,k=1}^N}{\Delta(z_1,...,z_N)}.
\ee
The main difference with the model \eqref{zour} considered in this work is the presence of derivatives instead of integral shifts. We observe that the following modification of \eqref{zvafa}
\be
\label{matrixmodel}
\int_{\H_N}\exp\tr\left(ZM-\frac 2\epsilon\cosh M\right)\frac{\Delta\left(\e^M\right)\d M}{\Delta(M)}=
2^N\pi^{\frac{N(N-1)}2}\frac{\det\left(\K_{z_j+k-1}\left(\frac 2\epsilon\right)\right)_{j,k=1}^N}{\Delta(z_1,...,z_N)}
\ee
(which coincides with \eqref{zvafa} for $N=1$ only) produces a result which is closer to the model \eqref{zour} under consideration in this work; indeed, from \eqref{fbessel} and the formula $\K_\nu(\zeta)=\frac \pi 2 \i^{\nu+1}\H^{(1)}_\nu(\i\zeta)$, one concludes that the transformations $\epsilon\mapsto \i\epsilon$, $z_j\mapsto z_j-\frac 1 2$ essentially convert \eqref{zour} to \eqref{matrixmodel}. As above, $\Delta(A)$ denotes the discriminant of the characteristic polynomial of the matrix $A$. The equality in \eqref{matrixmodel} is proven by the same arguments above, noting that after the angular integration using the Harish--Chandra-Itzykson-Zuber formula the left side is written as
\be
\frac{\pi^{\frac{N(N-1)}2}}{N!}\frac{1}{\Delta(Z)}\int_{\R^N}\Delta\left(\e^X\right)\det\left(\e^{x_iz_j-\frac 2\epsilon\cosh x_i}\right)_{i,j=1}^N\d X
\ee
and now the equality follows again from \eqref{andreief}, this time with $\phi_i(x):=\e^{(i-1)x}$.

Finally let us note that \eqref{matrixmodel} admits the alternative expression
\be
\int_{\H_N^+}\exp\tr\left(Z\log M-\frac 1\epsilon (M+M^{-1})\right)\frac{\Delta\left(\log M\right)}{\Delta(M)}\frac{\d M}{\det M}
\ee
where $\H_N^+$ is the cone of positive definite $N\times N$ hermitian matrices.

\begin{remark}
This matrix model has been obtained by completely independent means in the recent paper \cite{Ale20} from the free-fermion description of the Gromov-Witten theory of $\P^1$ \cite{OkPaequi}. Moreover, in \cite{Ale20} it is shown that a simple modification also describes the stationary sector of the Gromov-Witten theory of $\P^1$ relative to one point.
\end{remark}

\paragraph*{Connection with the Charlier ensemble.}

Introduce a discrete measure
\be
\label{poisson}
\mu_a:=\sum_{n\geq 0}\frac{\e^{-a}a^n}{n!}\delta_{n+\frac 12}
\ee
supported on $\left\lbrace\frac 12,\frac 32,\frac 52,...\right\rbrace$; here $\delta_{\xi}$ is the Dirac delta measure supported at $\xi\in\R$ and $a>0$ is a parameter. The monic discrete orthogonal polynomials $\pi_\ell(x;a)=x^\ell+\cdots$ relative to the measure \eqref{poisson} are known to be the (suitably scaled) \emph{Charlier polynomials};
\begin{align*}
\pi_\ell(x;a):&=(-a)^\ell{_2F_0}\left(-\ell,\frac 12 -x;;-\frac 1a\right),
\\
\int_\R \pi_\ell(x;a)\pi_{\ell'}(x,a)\d\mu_a(x)&=\sum_{n\geq 0}\pi_\ell\left(n+\frac 12;a\right)\pi_{\ell'}\left(n+\frac 12;a\right)\frac{\e^{-a}a^n}{n!}=a^{\ell}\ell!\delta_{\ell,\ell'}.
\end{align*}

The following result concerning a scaling limit of these orthogonal polynomials has been communicated to us by P. Lazag.

\begin{lemma}[\cite{Lazag}]
\label{lemmalazag}
For all $\zeta\in\R$ and $\ell\in\mathbb{Z}$ we have
\be
\lim_{L\to +\infty}\frac{\pi_{L+\ell}\left(L+\zeta;\frac 1{L\epsilon^2}\right)}{\G(L+\zeta+\frac 12)}=\epsilon^{\zeta-\ell-\frac 12}\J_{\zeta-\ell-\frac 12}\left(\frac 2\epsilon\right),
\ee
where $\J_\nu(\zeta)$ is the Bessel function of the first kind.
\end{lemma}

Consider now a matrix model of $L\times L$ hermitian matrices with spectrum distributed according to the discrete measure \eqref{poisson} (\emph{Charlier ensemble}).  In particular, the probability distribution of the eigenvalues is given by
\be
\label{charlierensemble}
\frac 1{Z_{L,a}}\Delta^2(x_1,...,x_L)\d\mu_a^{\otimes L}(x_1,...,x_L)
,\qquad Z_{L,a}:=\int_{\R^L}\Delta^2(x_1,...,x_L)\d\mu_a(x_1)\cdots\d\mu_a(x_L).
\ee
According to general results \cite{BrezinHikamicharacteristic,deiftcharacteristic} the expectation value of a product of characteristic polynomials admits the following expression
\be
\label{expectation}
\left\langle\prod_{i=1}^N\det\left(u_i\1-M\right)\right\rangle_{L,a}=\frac{\det\left(\pi_{L+k-1}(u_j)\right)_{j,k=1}^N}{\Delta(u_1,...,u_N)}
\ee
in terms of the monic orthogonal polynomials $\pi_0,\pi_1,...$; here the expectation value is taken according to the distribution \eqref{charlierensemble}.

Combining \eqref{expectation} with Lemma \ref{lemmalazag} we obtain the following interpretation of \eqref{zour}.

\begin{proposition}
For all $N\geq 1$, $z_1,...,z_N\in\R$, and all $\epsilon>0$, we have the following scaling limit of the expectation value of the product of $N$ characteristic polynomials in the Charlier ensemble, as the size $L$ diverges;
\be\label{scaling}
\lim_{L\to+\infty}\frac{\left\langle\prod_{i=1}^N\det\left(\left(L-z_i\right)\1-M\right)\right\rangle_{L,a=\frac 1{L\epsilon^2}}}{\prod_{i=1}^N\G(L-z_i+\frac 12)}=
\frac{\det \left(\epsilon^{\frac 12 -z_j-k}\J_{\frac 12 -z_j-k}\left(\frac 2\epsilon\right)\right)_{j,k=1}^N}{\Delta(z_1,...,z_N)}
\ee
where the expectation value in the left side is taken according to the distribution \eqref{charlierensemble}, with the parameter $a$ being set to $a=\frac 1{L\epsilon^2}$.
\end{proposition}

Up to minor modifications, we recognize the model \eqref{zour} in the right side of \eqref{scaling}. More precisely, the asymptotic relation \eqref{inserting} below implies the following asymptotic relation
\be
\label{expectation2}
(-1)^{\frac {N(N-1)}2}\left(\frac\pi 2\right)^{\frac N2}\lim_{L\to+\infty}\frac{\left\langle\prod_{i=1}^N\det\left(\left(L-z_i\right)\1-M\right)\right\rangle_{L,a=\frac 1{L\epsilon^2}}}{\prod_{i=1}^N\G(L-z_i+\frac 12)\cos(\pi z_i)\left(\frac {z_i}\e\right)^{z_i}}\sim\Z_N(z_1,...,z_N)
\ee
as $z_j\to +\infty$, where $\Z_N(z_1,...,z_N)$ is defined in \eqref{zour}.

The relation \eqref{expectation2} may be compared with the appearance of the Kontsevich matrix model \eqref{zk} in the \emph{edge-of-the-spectrum} scaling limit of the expectation value of a product of characteristic polynomials \cite{Okounkov,BrezinHikamiedge,BeCauniversality}.

\begin{remark}
Since for all $n=0,1,2,...$
\be
\label{residue}
\res{z=n+\frac 12}\G\left(\frac 12-z\right)(\e^{\i\pi}a)^{z}=-\sqrt{\e^{\i\pi}a}\frac{a^n}{n!}
\ee
we observe that the partition function for the Charlier ensemble of $L\times L$ hermitian matrices introduced above admits the following alternative expression;
\be
\label{eguchiconnection}
\int_{\H_L\left(\mathcal{C}\right)}\exp\tr V(M)\d M,\qquad V(z):=\log\Gamma\left(\frac 12-z\right)+z\left(\log a+\i\pi\right)
\ee
where $\mathcal{C}$ is a contour from $\infty+\i\delta$ to $\infty-\i\delta$ ($\delta>0$) surrounding the positive real axis, and $\H_L(\mathcal{C})$ is the set of unitarily diagonalizable $L\times L$ matrices with spectrum on $\mathcal{C}$. This is seen as the integral \eqref{eguchiconnection} localizes at the simple poles of the Gamma function and is therefore expressed as a sum of the relative residues \eqref{residue}. 

Stirling approximation of the logarithm of the Gamma function in \eqref{eguchiconnection} seems to hint at a connection with the one-matrix model with logarithmic potential which was proposed in \cite{EguchiYang} (see also \cite{MarshakovNekrasovSW}) to describe the Gromov-Witten theory of $\P^1$ (and in particular, its stationary sector). Further speculation about this connection is beyond the scope of this work and is deferred to future investigation.
\end{remark}

\subsection{Outline of the proof of Thm. \ref{mainthm}}

In this section we describe the steps in the proof of Thm. \ref{mainthm} and the organization of the paper.

\paragraph*{Dubrovin-Yang-Zagier formul\ae.}
Crucial to the proof of this result are the explicit formul\ae\ for stationary GW invariants of $\P^1$, conjectured by Dubrovin and Yang in \cite{DuYaP1} and proven together with Zagier in \cite{DuYaZa2018} (and independently proven in \cite{Ma2017} within the framework of Topological Recursion). This result can be summarized as follows.

Introduce the $2\times 2$ matrix valued formal series
\be
\label{R0}
R(z;\epsilon):=\frac{\pi}{\epsilon\cos(\pi z)}\left(
\begin{array}{cc}
\J_{z-\frac 12}\left(\frac 2\epsilon\right) \\ \J_{z+\frac 12}\left(\frac 2\epsilon\right)
\end{array}
\right)
\left(
\begin{array}{cc}
\J_{-z-\frac 12}\left(\frac 2\epsilon\right) & \J_{-z+\frac 12}\left(\frac 2\epsilon\right)
\end{array}
\right)=\left(\begin{array}{cc} 1 & 0\\ 0&0\end{array}\right)+\O(z^{-1})
\ee
where $\J_\nu(z)$ are the Bessel functions of the first kind, identified with their formal expansions as $z\to+\infty$ \cite{AbSt1965}. Introduce also the expressions\footnote{We denote $\mathfrak{S}_n$ the symmetric group over $\{1,2,...,n\}$.}
\begin{align}
\label{S1}
S_1&=\frac 1\epsilon\left(\frac{\pi}{\epsilon\cos(\pi z)}\left(
\begin{array}{cc}
\J_{-z-\frac 12}\left(\frac 2\epsilon\right) & \J_{-z+\frac 12}\left(\frac 2\epsilon\right)\end{array}\right)\left(
\begin{array}{c}
\pa_z\J_{z-\frac 12}\left(\frac 2\epsilon\right)
\\
\pa_z\J_{z+\frac 12}\left(\frac 2\epsilon\right)
\end{array}
\right)+\log(\epsilon z)\right),
\\
\label{Sn}
\quad S_n&=-\frac 1n\sum_{\sigma\in\mathfrak{S}_n}\frac{\tr\left(R(z_{\s(1)};\epsilon)\cdots R(z_{\s(n)};\epsilon)\right)}{(z_{\s(1)}-z_{\s(2)})\cdots(z_{\s(n-1)}-z_{\s(n)})(z_{\s(n)}-z_{\s(1)})}-\frac{\delta_{n,2}}{(z_1-z_2)^2}
\end{align}
understood as formal series in $z_1^{-1},...,z_n^{-1}$; note that \eqref{Sn} is well defined in this sense, as it is regular along the diagonals $z_i=z_j$.

The main result conjectured in \cite{DuYaP1} and proven in \cite{DuYaZa2018} is that for the stationary GW invariants of $\P^1$ 
\be
\label{gwinvariants}
\left\langle\tau_{k_1}\cdots\tau_{k_n}\right\rangle_{\P^1,d}:=\sum_{g\geq 0}\epsilon^{2g-2}\int_{\left[\Mgn(\P^1;d)\right]}\psi_1^{k_1}\cdots\psi_n^{k_n}\ev_1^*\omega\cdots\ev_n^*\omega
\ee
entering the generating function \eqref{FP1}, we have an expression in terms of formal residues, namely for all $n\geq 1$, $k_1,...,k_n\geq 0$ the following identity holds true;
\be
\label{dubrovindi}
\left\langle\tau_{k_1}\cdots\tau_{k_n}\right\rangle_{\P^1,d}
=
(-1)^n\res{z_1=\infty}\cdots\res{z_n=\infty}S_n(z_1,...,z_n)\prod_{j=1}^n\frac{\epsilon^{k_j+1}z^{k_j+1}_j\d z_j}{(k_j+1)!}
.
\ee
In the case $n=1$, \eqref{dubrovindi} reproduces the explicit formula for one-point stationary GW invariants of $\P^1$ due to Pandharipande \cite{Pandha}.

The strategy of the proof of Thm. \ref{mainthm} can be summarized as follows; the logic is completely parallel to the one employed in \cite{BeCa2017,BeRuPenner,BeRuBGW}.

\begin{enumerate}
\item We identify the right hand side of \eqref{dubrovindi} as logarithmic derivatives of a tau function \emph{of isomonodromic type}. Indeed, as we shall recall in Sec. \ref{seclimiting} (e.g. see Lemma \ref{lemmanpoint}), logarithmic derivatives of arbitrary order of such tau functions can be expressed in terms of formal residues as in the right hand side of \eqref{dubrovindi}. 

\item We identify logarithmic derivatives of the tau function of isomonodromic type with the limiting coefficients in the aforementioned expansion of $\Z_N(z_1,...,z_N)$. To accomplish such identification, we also interpret $\Z_N(z_1,...,z_N)$, for any $N$, as the expansion \emph{in every sector} of a tau function of isomonodromic type (see Sec. \ref{secrhpn}). 
\end{enumerate}

We now outline this approach with more details.

\paragraph*{Tau functions of isomonodromic type. }
Isomonodromic tau function have been originally introduced in the context of isomonodromic deformations, where a matrix linear ODE $\frac{\pa}{\pa z}\Psi(z;\ss)=A(z;\ss)\Psi(z;\ss)$, with $A$ rational in $z$, is assumed to depend analytically on parameters $\ss$ in such a way that its \emph{generalized monodromy data} is constant in $\ss$ \cite{JiMiUe1981}. Location of poles of $A(z)$ are part of the parameters $\ss$.

The definition of isomonodromic tau function in loc. cit. was later rephrased (and generalized) in \cite{Be2010} for a general matrix Riemann-Hilbert problem (RHP)
\be
\G_+(z)=\G_-(z)J(z),\quad z\in\S,\qquad \G(\infty)=\1
\ee
posed on some piecewise smooth oriented contour $\Sigma$ in the complex $z$-plane, with a jump matrix $J$ defined on $\S$. Concretely, assume that $J(z)=J(z;\ss)$ depends analytically on parameters $\ss$, and is actually the restriction to $\S$ of one (or more) analytic function(s) of $z$. The \emph{Malgrange differential} $\Omega$ is then defined as the following one-form, in the open set in the parameter space $\{\ss\}$ where the RHP $\G_+(z;\ss)=\G_-(z;\ss)J(z;\ss)$ has a solution;
\be 
\label{malgrange}
\Omega:=\sum_j \Omega_j\d s_j, \quad \Omega_j:=\int_\Sigma\tr\left(\G^{-1}_-\frac{\pa\G_-}{\pa z}\frac{\pa J}{\pa s_j} J^{-1}\right)\frac{\d z}{2\pi\i}.
\ee
Remarkably, the differential of $\Omega$ depends on $J$ only, and not on the solution $\G$; in many cases, $\Omega$ (or some simple modification) is closed, and we can accordingly introduce the tau function $\tau(\ss)$ by
\be
\frac{\pa}{\pa s_j}\log\tau(\ss)=\Omega_j.
\ee
This recovers the original setting of \cite{JiMiUe1981} when the the RHP is associated with a matrix linear ODE with rational coefficients.

More concretely, in this work we shall consider the $2\times 2$ matrix version of the difference equation \eqref{difference}
\be
\label{differencematrix}
\Psi(z+1)=A(z)\Psi(z),\qquad A(z)=\left(\begin{array}{cc}
 \epsilon\left(z+\frac 12\right) & -1 \\ 1 & 0
\end{array}\right)
\ee
which has a unique formal solution in the form\footnote{We use the Pauli matrix $\s_3=\diag(1,-1)$.}
\be
\label{formalsolution}
(\1+\O(z^{-1}))\left(\frac{\epsilon z}{\e}\right)^{z\sigma_3}.
\ee
In Sec. \ref{secbare} we study the \emph{Stokes phenomenon} of \eqref{differencematrix}, i.e. we construct sectors in the $z$-plane, cut along $z<0$, and analytic solutions to \eqref{differencematrix} which have \emph{the same} asymptotic expansion, given by the formal solution \eqref{formalsolution}, in every sector. The connection matrices relating different analytic solutions constitute the \emph{monodromy data} of the difference equation \eqref{differencematrix}, and they essentially define jumps of a RHP $\G_{0+}(z)=\G_{0-}(z)J_0(z)$, see Sec. \ref{secbare}.

Then we shall fix some $N\geq 1$ and we add dependence on parameters $\z=(z_1,...,z_N)$ to the RHP constructed above, by dressing the jump matrices $J_0(z)$ as in \eqref{dressingrational}.
Associated to this RHP we then have a tau function, $\tau_N(\z)$, to be defined concretely in \eqref{taun} (see also Remark \ref{remarkconstant}).
Moreover, as it follows from the results of \cite{JiMi1980,BeCa2015} such tau function admits a representation in terms of the determinant of a \emph{characteristic matrix}. To state this result, let us introduce the following generalization of \eqref{zour}
\be
\wh \Z_N(z_1,...,z_N)=\frac{\det \left(\frac 1{\epsilon^{k-1}}(\G_0(z_j+k-1))_{1,1}\right)_{j,k=1}^N}{\Delta(z_1,...,z_N)}
\ee
where the piecewise analytic matrix $\G_0$ mentioned above is defined in Sec. \ref{secbare}; we stress that, by construction, the $(1,1)$-entry of $\G_0$ admits the \emph{same} asymptotic expansion \eqref{besselasy} in every sector, so that it is really a good generalization of \eqref{zour} to every sector of the complex $z$-plane.

\begin{shaded}
\begin{proposition}\label{propmain}
The tau function $\tau_N(\z)$, concretely defined in \eqref{taun}, coincides with $\wh \Z_N(\z)$;
\be
\tau_N(\z)=\wh \Z_N(\z).
\ee
\end{proposition}
\end{shaded}

Next, we consider a \emph{limiting} (in the sense $N\to\infty$) RHP, in terms of the standard Miwa times
\be
\label{standardMiwa}
T_k:=\frac 1k\left(\frac 1{z_1^k}+\cdots+\frac 1{z_N^k}\right), \qquad k\geq 1
\ee
related to the scaled Miwa times \eqref{scaledmiwa} by 
\be
\label{scaledvsstandardmiwa}
t_k=\frac{(k+1)!}{\epsilon^k}T_{k+1},\qquad k\geq 0.
\ee
As before, this problem is constructed by dressing the jump matrices $J_0(z)$, this time in terms of parameters $\T$ as prescribed by \eqref{dressinglimiting}. We also associate a tau function $\tau(\T)$ to this problem, see \eqref{taulimiting}.

\begin{proposition}
\label{propfinal}
Logarithmic derivatives of the tau function $\tau(\T)$ of \eqref{taulimiting} are expressed in terms of the stationary GW invariants of $\P^1$, see \eqref{gwinvariants};
\be
\frac{\pa^n\log\tau(\T)}{\pa T_{\ell_1}\cdots\pa T_{\ell_n}}=\frac{\ell_1!\cdots\ell_n!}{\epsilon^{\ell_1+\cdots+\ell_n-n}}
\left\langle\tau_{k_1}\cdots\tau_{k_n}\right\rangle_{\P^1,d}.
\ee
\end{proposition}

\paragraph*{Acknowledgements.}
This project has received funding from the European Union's H2020 research and innovation programme under the Marie Sk\l owdoska-Curie grant No. 778010 {\em  IPaDEGAN}.
The work of M.B. was supported in part by the Natural Sciences and Engineering Research Council of Canada (NSERC) grant RGPIN-2016-06660.

\section{Asymptotic analysis of the matrix difference equation}\label{secbare}

In this section we study asymptotics of solutions to the difference equation \eqref{difference}, so to encode its general solution in a $2\times 2$ matrix solution of \eqref{differencematrix}, piecewise analytic in suitable sectors, and having the same asymptotic expansion \eqref{formalsolution} in every sector.

From now on we omit the dependence on the parameter $\epsilon>0$, in the interest of clarity.

Solutions to the difference equation \eqref{difference} can be expressed by Mellin contour integrals; in particular we choose
\begin{align}
\nonumber
f(z)&:=\frac 1{\sqrt{2\pi\epsilon}}\int_{C_1}\exp\left(\frac 1\epsilon\left(x-\frac 1x\right)-\left(z+\frac 32\right)\log x\right) \d x,
\\
\label{fi}
g(z)&:=\frac 1{\i\sqrt{2\pi\epsilon}}\int_{C_2}\exp\left(\frac 1\epsilon\left(x-\frac 1x\right)-\left(z+\frac 32\right)\log x\right) \d x,
\end{align}
where $C_1,C_2$ are contours in the $x$-plane with a branch cut along $x<0$, $|\arg x|<\pi$, for the definition of $\log x$. More precisely

\begin{itemize}
\item $C_1$ starts from $0$ with $|\arg x|<\frac\pi 2$ and arrives at $\infty$ with $\frac \pi 2<\arg x<\pi$, and
\item $C_2$ starts from $\infty$ with $-\pi<\arg x<-\frac\pi 2$ and arrives at $\infty$ with $\frac \pi 2<\arg x<\pi$.
\end{itemize}

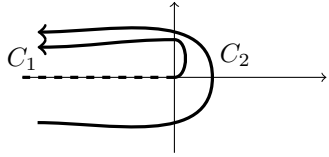
\begin{figure}[htbp]
\centering
\hspace{1cm}
\begin{tikzpicture}

\draw[dashed, very thick] (0,0) -- (-2,0);

\draw[->] (-2,0) -- (2,0);
\draw[->] (0,-1) -- (0,1);

\draw[->,very thick] (0,0) to[out=0,in=0] (0,.5) to[out=180, in=0] (-1.8,.4);
\draw[<-,very thick] (-1.8,.6) to[in=90, out=0] (.5,0) to[out=-90, in=0] (-1.8,-.6);

\node at (.8,.3) {$C_2$};
\node at (-2,.25) {$C_1$};
\end{tikzpicture}
\caption{Contours $C_1,C_2$ in the $x$-plane; the dashed line represents the branch cut along $x<0$ for the definition of $\log x$ in the integrand of\eqref{fi}.}
\label{figurecontour}
\end{figure}

These contours are depicted in Fig. \ref{figurecontour}. 

\begin{remark}
\label{rembessel}
$g$ can be expressed in terms of the Bessel function of first kind \cite{AbSt1965}
\be
g(z)=\sqrt{\frac {2\pi}\epsilon} \J_{z+\frac 12}\left(\frac 2\epsilon\right)
\ee
while $f$ can be expressed in terms of the Bessel function of first and second kind, or equivalently in terms of the Hankel function $\H^{(1)}$
\be
\label{fbessel}
f(z)=
\sqrt{\frac \pi{2\epsilon}}\left(\i \J_{z+\frac 12}\left(\frac 2\epsilon\right)-\Y_{z+\frac 12}\left(\frac 2\epsilon\right)\right)=\i\sqrt{\frac\pi {2\epsilon}}\H^{(1)}_{z+\frac 12}\left(\frac 2\epsilon\right).
\ee
Note that the $z$-dependence is in the order of the Bessel functions.
\end{remark}

\begin{lemma}
\label{lemmaasy}
The following asymptotic relations hold:
\begin{enumerate}
\item $f(z)\sim\left(\frac{\epsilon z}\e\right)^z(1+\O(z^{-1}))$, as $z\to\infty$ within $|\arg z|<\frac \pi 2-\delta$, for all $\delta>0$.
\item $g(z-1)\sim\left(\frac{\epsilon z}\e\right)^{-z}(1+\O(z^{-1}))$, as $z\to\infty$ within $|\arg z|\leq\pi-\delta$, for all $\delta>0$.
\end{enumerate}
\end{lemma}
\noindent The proof is based on the steepest descent method; we defer it to App. \ref{appsteepest}.

Let us fix angles $\a_1,...,\a_4$ satisfying
\be
\label{alpha}
-\pi<\a_1<-\frac \pi 2<\a_2<0<\a_3<\frac \pi 2<\a_4<\pi
\ee
and corresponding sectors in the $z$-plane, with a branch cut along $z<0$, $|\arg z|<\pi$;
\be
\SS_1:=\{-\pi<\arg z<\a_1\}, \quad \SS_j:=\{\a_{j-1}<\arg z<\a_j\} \ (j=2,3,4),\quad \SS_5:=\{\a_4<\arg z<\pi\}.
\ee

Define a piecewise analytic $2\times 2$ matrix $\Psi_0=\Psi_0(z)$ as
\be
\Psi_0(z):=\begin{cases}
\left(
\begin{array}{cc}
\e^{-\i\pi z}g(-z-1) & -\e^{\i\pi z}f(-z-1)
\\
-\e^{-\i\pi z}g(-z) & \e^{\i\pi z}f(-z)
\end{array}
\right),
& z\in \SS_1
\\
\\
\left(
\begin{array}{cc}
\frac 1{2\cos(\pi z)}g(-z-1) & g(z)
\\
-\frac 1{2\cos(\pi z)}g(-z) & g(z-1)
\end{array}
\right),
& z\in \SS_2
\\
\\
\left(
\begin{array}{cc}
f(z) & g(z) \\  f(z-1) & g(z-1)
\end{array}
\right),
& z\in \SS_3
\\
\\
\left(
\begin{array}{cc}
\frac 1{2\cos (\pi z)}g(-z-1) & g(z)
\\
-\frac 1{2\cos (\pi z)}g(-z) & g(z-1)
\end{array}
\right),
& z\in \SS_4
\\
\\
\left(
\begin{array}{cc}
\e^{\i\pi z}g(-z-1) & -\e^{-\i\pi z}f(-z-1)
\\
-\e^{\i\pi z}g(-z) &  \e^{-\i\pi z}f(-z)
\end{array}
\right),
& z\in \SS_5
\end{cases}
\ee
and define also
\be
\label{Psi}
\G_0(z):=\Psi_0(z)\left(\frac{\epsilon z}\e\right)^{-z\s_3}.
\ee

\begin{proposition}
\label{propbare}
The following statements hold  in all sectors $\SS_1,...,\SS_5$;
\begin{enumerate}
\item  The matrix $\Psi_0(z)$ solves the matrix difference equation \eqref{differencematrix}, and
\item The matrix $\G_0(z)$ admits an asymptotic expansion $\G_0(z)\sim \1+\O(z^{-1})$.
\end{enumerate}
\end{proposition}

\noindent{\bf Proof. }

\begin{enumerate}
\item Integrating by parts, we have ($i=1,2$)
\begin{align*}
0&=\int_{C_i}\pa_x\left(\e^{\frac 1\epsilon (x-\frac 1x)-(z+\frac 12)\log x}\right)\d x=\int_{C_i}\left(1+\frac 1{x^2}-\frac{z+\frac 12}x\right)\e^{\frac 1\epsilon (x-\frac 1x)-(z+\frac 12)\log x}\d x
\\
&=\int_{C_i}\left(\e^{\frac 1\epsilon (x-\frac 1x)-(z+\frac 12)\log x}+\e^{\frac 1\epsilon (x-\frac 1x)-(z+2+\frac 12)\log x}-\epsilon\left(z+\frac 12\right)\e^{\frac 1\epsilon (x-\frac 1x)-(z+1+\frac 12)\log x}\right)\d x
\end{align*}
which implies 
\be
f(z-1)+f(z+1)-\epsilon\left(z+\frac 12\right)f(z)=0=g(z-1)+g(z+1)-\epsilon\left(z+\frac 12\right)g(z).
\ee
Therefore the statement is true for the sector $\SS_3$. The statement in the remaining sectors is obtained noting that if $p(z)$ is any anti-periodic function $p(z+1)=-p(z)$, then $\wt f(z):=p(z)f(-z-1)$ and $\wt g(z):=p(z)g(-z-1)$ solve the same difference equation;
\be
\wt f(z-1)+\wt f(z+1)-\epsilon\left(z+\frac 12\right)\wt f(z)=0=\wt g(z-1)+\wt g(z+1)-\epsilon\left(z+\frac 12\right)\wt g(z).
\ee
\item In the sector $\SS_3$ the statement follows directly from Lemma \ref{lemmaasy}. For the sector $\SS_1$ we exploit the fact that $f,g$ are entire function and note that $0<\arg(\e^{\i\pi}z)<\frac \pi 2$, due to \eqref{alpha}, so that can apply Lemma \ref{lemmaasy} as
\begin{align*}
\e^{\i\pi z}f(-z)=\e^{\i\pi z}f(\e^{\i\pi}z)&\sim\e^{\i\pi z}\left(\frac{\epsilon\e^{\i\pi}z}{\e}\right)^{-z}(1+\O(z^{-1}))= \left(\frac{\epsilon z}{\e}\right)^{-z}(1+\O(z^{-1}))
\\
\e^{-\i\pi z}g(-z-1)=\e^{-\i\pi z}g(\e^{\i\pi}z-1)&\sim\e^{-\i\pi z}\left(\frac{\epsilon\e^{\i\pi}z}{\e}\right)^{z}(1+\O(z^{-1}))= \left(\frac{\epsilon z}{\e}\right)^{z}(1+\O(z^{-1})).
\end{align*}
The statement is proven likewise in the sectors $\SS_2,\SS_4,\SS_5$.
\QED
\end{enumerate}

Denote
\be
\S:=\e^{\i\a_1}\R_+\cup\cdots\cup \e^{\i\a_4}\R_+\cup \R_-
\ee
(rays oriented outwards) so that $\G,\Psi$ are analytic for $z\in\C\setminus\S=\SS_1\cup\cdots\cup \SS_5$

\begin{figure}[htbp]
\centering
\hspace{1cm}
\begin{tikzpicture}

\draw[->] (-2,0) -- (2,0);
\draw[->] (0,-2) -- (0,2);

\node at (-.9,-.5) {$\SS_1$};
\node at (.2,-1) {$\SS_2$};
\node at (1,.2) {$\SS_3$};
\node at (.2,1) {$\SS_4$};
\node at (-.9,.5) {$\SS_5$};

\begin{scope}[rotate=-120]
\node at (1.5,.25) {$+$};
\node at (1.5,-.25) {$-$};
\draw[->,very thick] (0,0) -- (1.5,0);
\draw[very thick] (1.5,0) -- (1.7,0);
\end{scope}

\begin{scope}[rotate=120]
\node at (1.5,.25) {$+$};
\node at (1.5,-.25) {$-$};
\draw[->,very thick] (0,0) -- (1.5,0);
\draw[very thick] (1.5,0) -- (1.7,0);
\end{scope}

\begin{scope}[rotate=-45]
\node at (1.5,.25) {$+$};
\node at (1.5,-.2) {$-$};
\draw[->,very thick] (0,0) -- (1.5,0);
\draw[very thick] (1.5,0) -- (1.7,0);
\end{scope}

\begin{scope}[rotate=+45]
\node at (1.5,.25) {$+$};
\node at (1.5,-.25) {$-$};
\draw[->,very thick] (0,0) -- (1.5,0);
\draw[very thick] (1.5,0) -- (1.7,0);
\end{scope}

\begin{scope}[rotate=180]
\node at (1.5,.2) {$+$};
\node at (1.5,-.15) {$-$};
\draw[->,very thick] (0,0) -- (1.5,0);
\draw[very thick] (1.5,0) -- (1.7,0);
\end{scope}

\end{tikzpicture}
\caption{Contour $\Sigma$, sectors $\SS_1,...,\SS_5$, and notation for the boundary values.}
\label{fig1}
\end{figure}
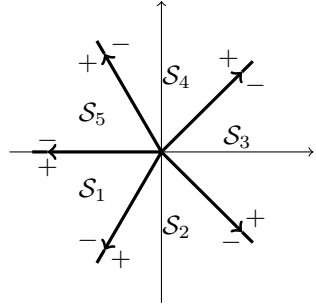

\begin{lemma}$\Psi_0(z)$ satisfies the jump condition
\be
\Psi_{0+}(z)=\Psi_{0-}(z)\wh J_0(z)
\ee
where the boundary values are taken according to the orientation of $\S$ (see Fig. \ref{fig1}) and the matrix $\wh J_0(z)$ is defined on $\S$ by
\be
\wh J_0(z)=\begin{cases}
\wh J_0^{(1)}(z)=
\left(
\begin{array}{cc}
\frac 1{1+q^{-1}}& \i q \\ 0 & 1+q^{-1}
\end{array}
\right), & z\in\e^{\i\a_1}\R_+
\\
\\
\wh J_0^{(2)}(z)=
\left( 
\begin{array}{cc}
1 & 0 \\ \frac \i{1+q} & 1
\end{array}
\right), & z\in\e^{\i\a_2}\R_+
\\
\\
\wh J_0^{(3)}(z)=
\left(
\begin{array}{cc}
1 & 0 \\ -\frac \i{1+q} & 1
\end{array}
\right), & z\in\e^{\i\a_3}\R_+
\\
\\
\wh J_0^{(4)}(z)=
\left(
\begin{array}{cc}
1+q & -\i \\ 0& \frac 1{1+q}
\end{array}
\right), & z\in\e^{\i\a_4}\R_+
\\
\\
\wh J_0^{(5)}(z)=
q^{-\s_3}, & z\in\R_-
\end{cases}
\ee
where we denote
\be
q:=\e^{2\pi\i z}.
\ee
\end{lemma}

\noindent{\bf Proof.}
It is a computation based on the identity
\be
g(-z-1)=2\cos(\pi z)f(z)-\i\e^{-\i\pi z}g(z),
\ee
which can be proven by performing the change of variable $x\mapsto -\frac 1x$ in the integral defining \eqref{fi} and applying the Cauchy theorem. Alternatively, in view of Rem. \ref{rembessel}, this identity follows from the known relation
\be
\label{knownidentity}
\J_{-\nu}(\zeta)=\i\sin(\pi\nu) \H_\nu^{(1)}(\zeta)+\e^{-\i\pi\nu}\J_\nu(\zeta)
\ee
of Hankel and Bessel functions \cite{AbSt1965}.
\QED

It follows that
\be
\G_{0+}(z)=\G_{0-}(z)J_0(z), \qquad J_0(z):=\left(\frac{\epsilon z_-}\e\right)^{z\s_3}\wh J_0(z)\left(\frac{\epsilon z_+}\e\right)^{-z\s_3}
\ee
where the notation for the boundary values in the definition of $J_0$ is relevant only along $z<0$.

The jump matrices $\wh J_0(z),J_0(z)$ satisfy the following properties.
\begin{enumerate}
\item $J_0^{(5)}(z)\equiv\1$, hence $\G_0$ extends analytically across $y<0$.
\item $J_0$ is exponentially close to the identity as $z\to\infty$, i.e. $J_0(z)=\1+\O\left(z^{-\infty}\right)$ as $z$ approaches $\infty$ along any of the rays $\e^{\i\a_j}\R_+$, $j=1,2,3,4$.
\item The no-monodromy condition $\wh J^{(1)}_0(z)\cdots\wh J^{(5)}(z)=\1$ holds true.
\item The jump matrices have unit determinant, $\det\wh J^{(i)}_0(z)\equiv 1$, $i=1,...,5$, $\det J^{(i)}(z)\equiv 1$, $i=1,...,4$.
\end{enumerate}

\begin{lemma}
\label{lemmadet}
We have $\det\Psi_0(z)\equiv 1\equiv\det\G_0(z)$ identically in $z,\epsilon$.
\end{lemma}

\noindent{\bf Proof.}
As $\det\wh J_0(z)$ is identically $1$ on $\Sigma$, we infer that $\Delta(z):=\det\Psi_0(z)$ is an entire function of $z$. Moreover, $\Delta$ is periodic, $\Delta(z+1)=\Delta(z)$, as it follows from \eqref{differencematrix}. Hence, $\Delta(z)\equiv 1$ everywhere.
\QED

\begin{remark}
The general results of Birkhoff \cite{Birkhoff} concerning existence of solutions to linear difference equations with appropriate asymptotics given by formal solutions to the same equation cannot be applied directly to \eqref{differencematrix}.
Indeed, in loc. cit. the general linear difference equation $\Psi(z+1)=A(z)\Psi(z)$ is analyzed assuming that $A(z)$ is a rational function of $z$ whose leading term in the expansion at $z=\infty$ has distinct \emph{nonzero} eigenvalues.
\end{remark}

\section{Matrix model and tau function: proof of Prop. \ref{propmain}}\label{secrhpn}

Let us denote $\S':=\e^{\i\a_1}\R_+\cup\cdots\cup\e^{\i\a_4}\R_+$.

Fix $N\geq 0$, points $\z=(z_1,...,z_N)$ in the complex plane, $|\arg z_j|<\pi$; by the freedom in the choice of the angles $\a_i$, compare with \eqref{alpha}, we can assume that $z_1,...,z_N\in\C\setminus\S'$.
Associated with this data, introduce the jump matrix $J_N(z;\z):\S'\to\SL_2(\C)$ by
\be
\label{dressingrational}
J_N(z;\z):=D^{-1}_N(z;\z)J_0(z)D_N(z;\z),\qquad D_N(z;\z):=\left(
\begin{array}{cc}
1 &0\\ 0&\prod_{j=1}^N(1-\frac z{z_j})
\end{array}\right).
\ee

\begin{shaded}
\begin{problem}
\label{rhprational}
Find a $2\times 2$ matrix $\G_N(z;\z)$, analytic in every sector of $\C\setminus\S'$, satisfying the following jump condition along $\S'$
\be
\G_{N+}(z;\z)=\G_{N-}(z;\z)J_N(z;\z),
\ee
and the following boundary condition at infinity
\be
\G_N(z;\z)\sim \1+\O(z^{-1}).
\ee
\end{problem}
\end{shaded}

\begin{remark}
As in Lemma \eqref{lemmadet} it can be shown that $\det\G_N(z;\z)\equiv 1$ identically in $z$, whenever $\G_N(z;\z)$ exists. Hence, the solution to the RHP \ref{rhprational} is unique, if it exists.
\end{remark}

Introduce the \emph{Jimbo-Miwa-Ueno differential} \cite{JiMiUe1981}
\be
\label{jmun}
\Omega_N=\sum_{j=1}^N\Omega_{N,j}\d z_j,\qquad \Omega_{N,j}:=\res{z=z_j}\tr\left(\G_N^{-1}\frac{\pa\G_N}{\pa z}\frac{\pa D_N}{\pa z_j}D^{-1}_N\right)\d z
\ee
and the \emph{Malgrange differential} \cite{Be2010}
\be
\label{malgrangen}
\wh\Omega_N=\sum_{j=1}^N\wh\Omega_{N,j}\d z_j, \qquad \wh\Omega_{N,j}:=\int_{\S'}\tr\left(\G_{N-}^{-1}\frac{\pa\G_{N-}}{\pa z}\frac{\pa J_N}{\pa z_j}J^{-1}_N\right)\frac{\d z}{2\pi\i}.
\ee

\begin{lemma}
\label{lemmamalgrangevsjmurhpn}
The following identity holds true;
\be
\Omega_N-\wh \Omega_N=\eta_N
\ee
where
\be
\label{eta}
\eta_N=\sum_{j=1}^N\eta_{N,j}\d z_j,\qquad \eta_{N,j}:=\int_{\S'}\tr\left(J_N^{-1}\frac{\pa J_N}{\pa z}\frac{\pa D_N}{\pa z_j}D^{-1}_N\right)\frac{\d z}{2\pi\i}.
\ee
\end{lemma}

\noindent{\bf Proof.}
Let us denote $\pa_j:=\frac{\pa}{\pa z_j}$, $':=\frac{\pa}{\pa z}$. From \eqref{dressingrational} we have $\pa_jJ_N=[J_N,\pa_j D_ND^{-1}_N]$ hence, exploiting the cyclic property of the trace
\be
\tr\left(\G_{N-}^{-1}\G'_{N-}\pa_jJ_NJ_N^{-1}\right)=\tr\left(\left(J_N^{-1}\G_{N-}^{-1}\G'_{N-}J_N-\G_{N-}^{-1}\G'_{N-}\right)\pa_jD_ND^{-1}_N\right)
\ee
and noting $\G'_{N+}=\G'_{N-}J_N+\G_{N-}J_N'$ we can rewrite the last expression as
\be
\tr\left(\left(\left[\G_N^{-1}\G'_N\right]^+_--J_N^{-1}J_N'\right)\pa_jD_ND^{-1}_N\right)=\tr\left(\left[\G_N^{-1}\G'_N\pa_jD_ND^{-1}_N\right]^+_-\right)-\tr\left(J_N^{-1}J_N'\pa_jD_ND^{-1}_N\right)
\ee
where $[f]^+_-:=f_+-f_-$. By Cauchy's theorem 
\be
\int_{\S'}\tr[\G_N^{-1}\G'_N\pa_jD_ND^{-1}_N]^+_-\frac{\d z}{2\pi\i}=\sum_{j=1}^N\res{z=z_j}\tr\left(\G_N^{-1}\G'_N\pa_jD_ND^{-1}_N\right)\d z
\ee
and the proof is complete.
\QED

The following was proven originally in \cite{JiMiUe1981}, in the slightly different context of isomonodromic deformations of a matrix linear ODE with rational coefficients. However, the proof applies equally well here.

\begin{theorem}
\label{lemmaomegajmuclosedrhpn}
The Jimbo-Miwa-Ueno differential is closed;
\be
\frac{\pa}{\pa z_j}\Omega_{N,k}=\frac{\pa}{\pa z_k}\Omega_{N,j}.
\ee
\end{theorem}

Hence we introduce the tau function $\tau_N(\z)$ as
\be
\label{taun}
\Omega_{N,j}=\frac{\pa}{\pa z_j}\log\tau_N(\z).
\ee

\begin{remark}\label{remarkconstant}
The tau function $\tau_n$ is  defined by equation \eqref{taun} only up to a  $\z$--independent multiplicative factor. In particular, Proposition \ref{propmain} is equivalent to the variational identity $\frac{\pa}{\pa z_j}\log \wh\Z_N(\z)=\Omega_{N,j}$, with $\Omega_{N,j}$ defined in \eqref{jmun}.
\end{remark}

From the theory of \emph{Schlesinger transformations} \cite{JiMi1980,BeCa2015}, we know that a tau function related to a rational dressing of jump matrices like \eqref{dressingrational}, admits an explicit expression in terms of a finite size determinant. We recall this  result applied to our setting.

Introduce the $N\times N$ \emph{characteristic matrix} $G_N(\z)$, with entries
\be
\label{characteristicmatrix}
(G_N(\z))_{j,k}=-\res{z=\infty}\left(\G^{-1}_0(z_j)\G_0(z)\right)_{2,2}\frac{z^{k-1}\d z}{z-z_j},\qquad 1\leq j,k\leq N.
\ee

The following result follows from \cite[App. B, Thm. 2.2]{BeCa2015}, hence we omit the proof.

\begin{theorem}
We have
\be
\Omega_{N,j}=\frac{\pa}{\pa z_j}\log\frac{\det G_N(\z)}{\prod_{1\leq a<b\leq N}(z_b-z_a)}
\ee
where the characteristic matrix $G_N(\z)$ is defined in \eqref{characteristicmatrix}.
\end{theorem}

Finally, Prop. \ref{propmain} follows from the following computation of the determinant of the characteristic matrix \eqref{characteristicmatrix}.

\begin{proposition}
We have
\be
\det G_N(\z)=\det \left(\frac 1{\epsilon^{k-1}}(\G_0(z_j+k-1))_{1,1}\right)_{j,k=1}^N.
\ee
\end{proposition}

\noindent{\bf Proof.}
Introduce functions $a(z),b(z)$, analytic in every sector $\SS_1,...,\SS_5$, according to
\be
\label{ab}
\Psi_0(z)=\left(
\begin{array}{cc}
a(z) & b(z) \\ a(z-1) & b(z-1)
\end{array}
\right)
\ee
so that the entries \eqref{characteristicmatrix} of the characteristic matrix are found as
\be
\label{lllast}
\left(\frac{\epsilon z_j}{\e}\right)^{-z_j}\left(\frac{\epsilon z}{\e}\right)^{z}\frac{\left(\begin{array}{cc}-a(z_j-1)&a(z_j)\end{array}\right)\left(\begin{array}{cc}b(z) \\ b(z-1)\end{array}\right)}{z-z_j}=\sum_{k=1}^N(G_{N}(\z))_{j,k}z^{-k}+\O(z^{-N-1})
\ee
where we use $\det\G_0(z)\equiv 1$ from Lemma \ref{lemmadet}. Introducing the matrix
\be
H(z;z_j)=\left(
\begin{array}{cc}
a(z_j) & b(z) \\ a(z_j-1) & b(z-1)
\end{array}
\right)
\ee
we rewrite \eqref{lllast} as
\be
\label{llast}
\left(\frac{\epsilon z_j}{\e}\right)^{-z_j}\left(\frac{\epsilon z}{\e}\right)^{z}\frac{\det H(z;z_j)}{z-z_j}=\sum_{k=1}^N(G_{N}(\z))_{j,k}z^{-k}+\O(z^{-N-1}).
\ee
Recalling the difference equation \eqref{differencematrix} we have
\be
\left(
\begin{array}{cc}
a(z_j+1) & b(z+1)+\epsilon(z_j-z)b(z) \\ a(z_j-1) & b(z-1)
\end{array}
\right)=\left(
\begin{array}{cc}
\epsilon\left(z_j+\frac 12\right) & -1 \\ 1 & 0
\end{array}
\right)
\left(
\begin{array}{cc}
a(z_j) & b(z) \\ a(z_j-1) & b(z_j)
\end{array}
\right)
\ee
hence we get
\be
\det H(z+1;z_j+1)+\epsilon(z-z_j)a(z_j-1)b(z)=\det H(z;z_j)
\ee
from which we obtain
\be
\det H(z+N;z_j+N)+\epsilon(z-z_j)\sum_{\ell=1}^Nb(z+\ell)a(z_j+\ell-1)=\det H(z;z_j).
\ee
Finally, from $\left(\frac{\epsilon z}{\e}\right)^{z}b(z+\ell)=\frac 1{(\epsilon z)^\ell}\wh b_\ell(z)$, $\wh b_\ell(z)=1+\O(z^{-1})$ we rewrite \eqref{llast} as
\be
\left(\frac{\epsilon z_j}{\e}\right)^{-z_j}\sum_{\ell=1}^N\frac{a(z_j+\ell-1)\wh b_\ell(z)}{\epsilon^{\ell-1} z^\ell}=\sum_{k=1}^N(G_{N}(\z))_{j,k}z^{-k}+\O(z^{-N-1})
\ee
and so
\be
\label{GGB}
G_N(\z)=\wh G_N(\z) B_N
\ee
where we write $\wh b_\ell(z)=1+\sum_{j\geq 1}\wh b_\ell^jz^{-j}$ and
\be
B_N:=\left(
\begin{array}{cccc}
1 & \wh b_1^1 & \cdots & \wh b_1^{N-2} \\
0 & 1 & \cdots & \wh b_2^{N-3} \\
\vdots &\vdots &\ddots &\vdots \\
0&0&\cdots & 1
\end{array}
\right), \quad \quad (\wh G_N(\z))_{j,k}=\left(\frac{\epsilon z_j}{\e}\right)^{-z_j}\frac{a(z_j+k-1)}{\epsilon^{k-1}}=
\frac{\left(\G_0(z_j+k-1)\right)_{1,1}}{\epsilon^{k-1}}
\ee
and the proof is complete by taking the determinant of identity \eqref{GGB}, as $\det B_N\equiv 1$.
\QED

This completes the proof of Prop. \ref{propmain}

\section{The limiting Riemann-Hilbert problem}\label{seclimiting}

For all $N\geq 0$, we have the identity
\be
\label{booooh}
D_N^{-1}=\left(
\begin{array}{cc}
1 & 0 \\ 0& \e^{\sum_{\ell\geq 1}T_\ell(\z) z^\ell}
\end{array}
\right), \qquad T_\ell(\z):=\frac 1\ell \sum_{j=1}^N z^{-\ell}_j.
\ee
This identity  is non-formal provided $\min_{j=1,...,N}|z_j|>|z|$. 

This prompts to introduce an independent set of times $T_1,T_2,...$, and\footnote{Here $\E_{22}$ is the elementary matrix $\left(\begin{array}{cc}0 & 0 \\ 0& 1\end{array}\right)$.}
\be
\label{dressinglimiting}
J(z;\T):=\e^{\vartheta(z;\T)\E_{22}}J_0(z)\e^{-\vartheta(z;\T)\E_{22}},\qquad \vartheta(z;\T):=\sum_{\ell\geq 1}T_\ell z^\ell
\ee
and to consider the following RHP.

\begin{shaded}
\begin{problem}
\label{rhplimiting}
Find a $2\times 2$ matrix $\G(z;\T)$, analytic in every sector of $\C\setminus\S'$, satisfying the following jump condition along $\S'$
\be
\G_+(z;\T)=\G_-(z;\T)J(z;\T),
\ee
and the following boundary condition at infinity
\be
\G(z;\T)\sim\1+\O(z^{-1}).
\ee
\end{problem}
\end{shaded}

\paragraph*{Analytic discussion of RHP \ref{rhplimiting}, and limit of Problem \ref{rhprational} and of the tau function.}

For the sake of definiteness, in the RHP \ref{rhplimiting} one must first assume that for some $K\geq 1$ we have $T_\ell=0$ whenever $\ell>K$. In principle, this assumption is in contradiction with the interpretation of the $T_\ell$'s as the standard Miwa variables of the $z_j$'s \eqref{standardMiwa}, appearing in \eqref{booooh}; this interpretation is relevant in order to regard the RHP \ref{rhplimiting} as an analytic limit of the RHP \ref{rhprational}. We now briefly address this issue.

Let us fix an arbitrary $K\geq 1$ and assume $T_\ell=0$ whenever $\ell>K$. Under the assumption that
\be
\label{argumentTK}
\Re t_K\e^{\i K\a_j}<0,\quad j=2,3,\qquad \Re t_K\e^{\i K\a_j}>0,\quad j=1,4
\ee
we conclude that $J(z;\T)=\1+\O(z^{-\infty})$ as $z\to\infty$ along any ray of $\S$. Hence, the solution to the RHP \ref{rhplimiting} exists and is unique for $\T = (T_1,...,T_K)$ in an open neighborhood of $\T=0$, with the argument of $T_K$ further restricted by \eqref{argumentTK}; it defines a matrix function $\G(z;\T)$, its specifications to each sector of the $z$-plane being holomorphic in $T_1,...,T_K$. Note that $\G(\T=0)=\G_0$ by construction.

In particular, this allows to introduce the Jimbo-Miwa-Ueno and the Malgrange differentials as above, see \eqref{jmun}-\eqref{malgrangen};
\begin{align}
\Omega&=\sum_{\ell=1}^K\Omega_\ell\d T_\ell, \qquad \Omega_\ell:=-\res{z=\infty}\tr\left(\G^{-1}\frac{\pa\G}{\pa z}\E_{22}\right)z^\ell\d z,
\\
\wh\Omega&=\sum_{\ell=1}^K\wh\Omega_\ell\d T_\ell, \qquad \wh \Omega_\ell:=\int_{\S'}\tr\left(\G^{-1}_-\frac{\pa\G_-}{\pa z}\frac{\pa J}{\pa T_\ell}J^{-1}\right)\frac {\d z}{2\pi\i}
.
\end{align}
Exactly as in Lemma \ref{lemmamalgrangevsjmurhpn}, we establish the relation
\be
\Omega=\wh\Omega+\eta,\qquad \eta=\sum_{\ell=1}^K\eta_\ell\d T_\ell, \quad \eta_\ell:=-\int_{\S'}\tr\left(J^{-1}\frac{\pa J}{\pa z}\E_{22}\right)\frac{z^\ell\d z}{2\pi\i}.
\ee
Moreover, it can also be proven that $\Omega$ is closed \cite{JiMiUe1981}
\be
\frac{\pa}{\pa T_j}\Omega_k=\frac{\pa}{\pa T_k}\Omega_j
\ee
and so we can introduce the tau function $\tau(\T)$ as
\be
\label{taulimiting}
\Omega_\ell=\frac{\pa}{\pa T_\ell}\log\tau(\T).
\ee
It follows that $\tau(\T)$ is an analytic function of $T_1,...,T_K$ in an open neighborhood of $\T=0$ with $\arg T_K$ restricted by \eqref{argumentTK}. 

The main goal now is to identify the Taylor expansion of $\tau(\T)$ at $\T=0$ with the formal limiting expansion of $\Z_N(z_1,...,z_N)$ as $N\to\infty$ in terms of the standard Miwa variables \eqref{standardMiwa}. This can be analytically achieved by the following argument. For fixed $\T=(T_1,...,T_K,0,...)$ introduce, for $N\geq 1$, the roots $\z^{(N)}=(z_1^{(N)},...,z_N^{(N)})$ of the Taylor polynomials of $\e^{\vartheta(z;\T)}$, i.e.
\be
\exp\left(T_1z+\cdots+T_Kz^K\right)=(z-z_1^{(N)})\cdots(z-z_N^{(N)})+\O(z^{N+1}).
\ee
Then one should check convergence (as $N\to\infty$, uniformly for $T_1,...,T_K$ in compact sets) in all $L^p$ norms of the jump matrices for the RHP \ref{rhprational}, defined in terms of $\z^{(N)}$, to those of the RHP \ref{rhplimiting}. Then standard perturbation analysis of RHP permits to deduce convergence of $\G_N(z;\z^{(N)})$ to $\G(z;T_1,...,T_K,0,...)$ and of tau functions $\tau_N(\z_N^{(N)})$ to $\tau(T_1,...,T_K,0,...)$, using the representations $\Omega_N=\wh\Omega_N-\eta_N$ and $\Omega=\wh\Omega-\eta$ for logarithmic derivatives of the tau functions. Similar convergence of logarithmic derivatives of the tau functions can be deduced then by the fact the latter admit expressions in terms of the solution $\G$ to the RHP only, see e.g. Lemma \ref{lemmanpoint}.

For the main purpose of this work we are mostly concerned with the formal aspects of RHP \ref{rhplimiting}, and the considerations above then play a minor role, so we refer to the detailed analysis for the Kontsevich-Witten tau function of \cite{BeCa2017}, which is essentially similar to the case under consideration in this work.

\section{Tau function and $\mathbb P^1$ Gromov-Witten invariants: proof of Prop. \ref{propfinal}}

We first consider one-point intersection numbers, $n=1$. To this end, applying definition \eqref{taulimiting}, using the notation of \eqref{ab} and denoting $':=\pa_z$, we compute
\be
\left.\frac{\pa}{\pa T_\ell}\log\tau(\T)\right|_{\T=0}=-\res{z=\infty}\left(\left(
\begin{array}{cc}
-a(z-1) & a(z)
\end{array}
\right)
\left(
\begin{array}{c}
b'(z) +b(z)\log (\epsilon z)\\ b'(z-1)+b(z-1)\log (\epsilon z)
\end{array}
\right)\right)z^\ell\d z
\ee
where we use the identity $\left(\frac{\epsilon z}\e\right)^{-z}\left(\left(\frac{\epsilon z}\e\right)^z\right)'=\log(\epsilon z)$. Since
\be
\det \Psi_0(z)=a(z)b(z-1)-a(z-1)b(z)\equiv 1
\ee
we can write
\be
\label{verylast}
\left.\frac{\pa}{\pa T_\ell}\log\tau(\T)\right|_{\T=0}=-\res{z=\infty}\left(\left(
\begin{array}{cc}
-a(z-1) & a(z)
\end{array}
\right)
\left(
\begin{array}{c}
b'(z)\\ b'(z-1)
\end{array}
\right)+\log (\epsilon z)\right)z^\ell\d z.
\ee
The formal residue is independent of the sector in which we let $z\to\infty$ by construction, as $\G_0(z)$ has the same asymptotic expansion in every sector. E.g. we can assume, using the definition of $\G_0(z)$ in the sector $\SS_3$, compare with \eqref{Psi}, that
\be
\label{inserting}
a(z)=f(z)=\i\sqrt{\frac{\pi}{2\epsilon}}\H^{(1)}_{z+\frac 12}\left(\frac 2\epsilon\right)\sim
\sqrt{\frac{\pi}{2\epsilon}}\frac 1{\cos(\pi z)}\J_{-z-\frac 12}\left(\frac 2\epsilon\right),\qquad b(z)=g(z)=\sqrt{\frac{2\pi}\epsilon}\J_{z+\frac 12}\left(\frac 2\epsilon\right)
\ee
where we use the Hankel function $\H^{(1)}_\nu(\zeta)=\J_\nu(\zeta)+\i\Y_\nu(\zeta)$, the identity
\be
\H^{(1)}_\nu(\zeta)=\frac{\i}{\sin(\nu\pi)}\left(\e^{-\nu\pi\i}\J_\nu(\zeta)-\J_{-\nu}(\zeta)\right)
\ee
compare with \eqref{knownidentity} \cite{AbSt1965}, and the fact that the term involving $J_{z+\frac 12}\left(\frac 2\epsilon\right)$ is sub-leading as $z\to+\infty$, hence inconsequential for the computation of the formal residue \eqref{verylast}. Inserting \eqref{inserting} in \eqref{verylast} we obtain
\begin{align}
&\left.\frac{\pa}{\pa T_\ell}\log\tau(\T)\right|_{\T=0}
\\
&\qquad=-\res{z=\infty}\left(\frac \pi{\epsilon \cos (\pi z)}\left(
\begin{array}{cc}
\J_{-z+\frac 12}\left(\frac 2 \epsilon\right) & \J_{-z-\frac 12}\left(\frac 2 \epsilon\right)
\end{array}
\right)
\left(
\begin{array}{c}
\pa_z\J_{z+\frac 12}\left(\frac 2 \epsilon\right)\\ \pa_z\J_{z-\frac 12}\left(\frac 2 \epsilon\right)
\end{array}
\right)+\log (\epsilon z)\right)z^\ell\d z
\\
&\qquad=-\res{z=\infty}\epsilon S_1(z)z^{\ell}\d z=\frac{\ell!}{\epsilon^{\ell-1}}\left\langle\tau_{\ell-1}\right\rangle_{\P^1,d}
\end{align}
where we used \eqref{S1} and \eqref{dubrovindi}. This proves Prop. \ref{propfinal} for $n=1$.

In order to proceed with higher order derivatives, we first note that we have a compatible system of ODEs of the form
\be
\frac{\pa\G}{\pa T_\ell}=M_\ell \G-z^\ell\G\E_{22}, \qquad \frac{\pa M_m}{\pa T_\ell}-\frac{\pa M_\ell}{\pa T_m}=[M_\ell,M_m]
\ee
where $M_\ell=M_\ell(z;\T)$ is a polynomial of degree $\ell$ in $z$;
\be
\label{M}
M_\ell(z):=\res{w=\infty}\frac{\G(w;\T)\E_{22}\G^{-1}(w;\T)}{w-z}{w}^\ell\d w=\res{w=\infty}\frac{U(w;\T)}{w-z}w^\ell\d w,\quad \ell\geq 1
\ee
where
\be
U(z;\T):=\G(z;\T)\E_{22}\G^{-1}(z;\T).
\ee
This fact follows by a standard application of the Liouville theorem. The matrix $\G\e^{\vartheta\E_{22}}$ is piecewise analytic in the complex $z$-plane and satisfies jump conditions independent of $\T$ along $\S'$. Hence the ratio $\frac{\pa}{\pa T_\ell}\left(\G\e^{\vartheta\E_{22}}\right)\left(\G\e^{\vartheta\E_{22}}\right)^{-1}=:M_\ell$ is analytic in $z$ everywhere and grows like a polynomial of degree $\ell$ at $z=\infty$. It follows that $M_\ell$ can be found as the polynomial part of the expansion at $z=\infty$, as in \eqref{M}.

Then we compute second derivatives of $\log\tau(\T)$, using the cyclic property of the trace and denoting $':=\pa_z$;
\begin{align*}
\frac{\pa}{\pa T_{\ell_2}}\frac{\pa}{\pa T_{\ell_1}}\log\tau(\T)&=-\res{z_1=\infty}\tr\frac{\pa}{\pa T_{\ell_2}}\left(\G^{-1}(z_1;\T)\G'(z_1;\T)\E_{22}\right)z^{\ell_1}_1\d z_1
\\
&=-\res{z_1=\infty}\tr\left(\G^{-1}(z_1;\T)M'_{\ell_2}(z_1;\T)\E_{22}\G(z_1;\T)\right)z^{\ell_1}_1\d z_1+\res{z_1=\infty}\left(\ell_2 z^{\ell_1+\ell_2-1}_1\right)\d z_1
\\
&=\res{z_1=\infty}\res{z_2=\infty}\frac{\tr\left(U(z_1;\T)U(z_2;\T)\right)-1}{(z_1-z_2)^2}z^{\ell_1}_1z^{\ell_2}\d z_1\d z_2.
\end{align*}

\begin{lemma}
\label{lemma0}
In the sense of asymptotic expansions at $z=\infty$, we have
\be
U(z;\T)\big|_{\T=0}=\s_1 R(z) \s_1,\qquad \s_1:=\left(\begin{array}{cc}
0&1 \\ 1&0
\end{array}\right)
\ee
where $R(z)=R(z;\epsilon)$ is given in \eqref{R0}.
\end{lemma}

\noindent{\bf Proof.}
Using the notation of \eqref{ab} we compute
\be
U(z;\T)\big|_{\T=0}=\G_0(z)\E_{22}\G_0^{-1}(z)=\left(
\begin{array}{c}
b(z) \\ b(z-1)
\end{array}
\right)
\left(\begin{array}{cc}-a(z-1) & a(z)\end{array}\right)
\ee
and so the proof is complete by comparing with \eqref{inserting}.
\QED

Comparing with \eqref{Sn} and \eqref{dubrovindi} for $n=2$ we conclude that Prop. \ref{propfinal} is also true for $n=2$.

To complete the proof of Prop. \ref{propfinal} we state the next lemma. We omit its proof as it is based on algebraic manipulations by induction that have appeared several times in the literature; e.g. we refer the reader to \cite{BeDuYa2016,BeRuPenner,BeRuBGW}.

\begin{lemma}
\label{lemmanpoint}
We have
\begin{align*}
&\frac{\pa^n\log\tau(\T)}{\pa T_{\ell_1}\cdots \pa T_{\ell_n}}
\\
&=(-1)^n\res{z_1=\infty}\cdots\res{z_n=\infty}\left(-\frac 1n \sum_{\sigma\in\mathfrak{S}_n}\frac{\tr\left(U(z_{\s(1)};\T)\cdots U(z_{\s(n)};\T)\right)}{(z_{\s(1)}-z_{\s(2)})\cdots (z_{\s(n)}-z_{\s(1)})}-\frac{\delta_{n,2}}{(z_1-z_2)^2}\right)\d z_1\cdots \d z_n.
\end{align*}
\end{lemma}

Finally, the proof of Prop. \ref{propfinal} is complete by setting $\T=0$, applying Lemma \ref{lemma0}, and comparing with \eqref{Sn} and \eqref{dubrovindi}.

\appendix \normalfont
\renewcommand{\theequation}{\Alph{section}.\arabic{equation}}

\section{Asymptotics for $f,g$: proof of Lemma \ref{lemmaasy}}\label{appsteepest}

It is convenient to introduce
\begin{align*}
\wh f(z)&:=\int_{C_1}\exp\left(\frac 1\epsilon\left(x-\frac 1x\right)-\left(z+\frac 32\right)\log x\right) \d x,
\\
\wh g(z)&:=\int_{C_2}\exp\left(\frac 1\epsilon\left(x-\frac 1x\right)-\left(z+\frac 32\right)\log x\right) \d x.
\end{align*}

\paragraph*{Asymptotics for $\wh g$.}
Let us write $\xi:=\epsilon\left(z+\frac 12\right)$ so that
\be
\wh g(z-1)
=\int_{C_2}\e^{\frac 1\epsilon\left(x-\frac 1x-\xi\log x\right)}\d x=
|\xi|\e^{-\frac \xi \epsilon \log|\xi|}\int_{C_2}\e^{\frac {|\xi|}\epsilon\left(x-\e^{\i\theta}\log x\right)}\e^{-\frac 1{\epsilon|\xi|x}}\d x
\ee
where $\xi=|\xi|\e^{\i\theta}$, $|\theta|<\pi$; in the second equality we performed the change of variable $x\mapsto x|\xi|$ and applied Cauchy theorem to deform the contour $|\xi|^{-1}C_2$ back to $C_2$. Since $C_2$ stays at a bounded distance from $x=0$, we can apply Fubini theorem and write
\be
\label{last}
\int_{C_2}\e^{\frac {|\xi|}\epsilon\left(x-\e^{\i\theta}\log x\right)}\e^{-\frac 1{\epsilon|\xi|x}}\d x=
\sum_{j\geq 0}\frac{(-1)^j}{j!\epsilon^j|\xi|^j}
\int_{C_2}\frac 1{x^j}\e^{\frac {|\xi|}\epsilon\left(x-\e^{\i\theta}\log x\right)}\d x.
\ee

We study each integral in the series in right hand side of \eqref{last} by the steepest descent method. The phase is $\varphi(x):=x-\e^{\i\theta}\log x$, which has one saddle point at $x=\e^{\i\theta}$. Expanding $\varphi(x)=\e^{\i\theta}(1-\i\theta)+\frac{\e^{-\i\theta}}2(x-\e^{\i\theta})^2+\O((x-\e^{\i\theta})^3)$ we see that the steepest descent direction is $\frac{\pi+\theta}2$.

\begin{figure}[htbp]
\centering
\hspace{1cm}
\begin{subfigure}{.22\textwidth}
\includegraphics[scale=.2]{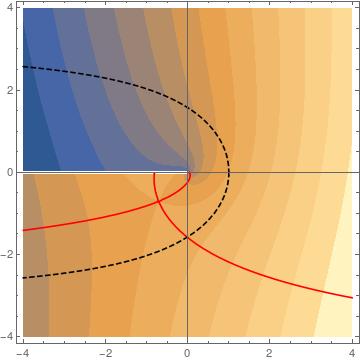}
\caption*{$\theta=-\frac {3\pi}4$}
\end{subfigure}
\begin{subfigure}{.22\textwidth}
\includegraphics[scale=.2]{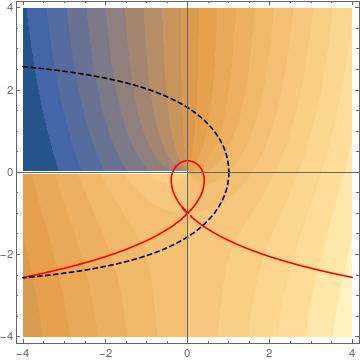}
\caption*{$\theta=-\frac \pi 2 \pi$}
\end{subfigure}
\begin{subfigure}{.22\textwidth}
\includegraphics[scale=.2]{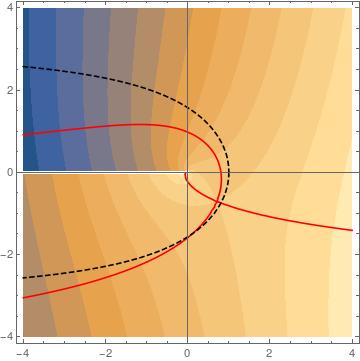}
\caption*{$\theta=-\frac \pi 4$}
\end{subfigure}
\begin{subfigure}{.22\textwidth}
\includegraphics[scale=.2]{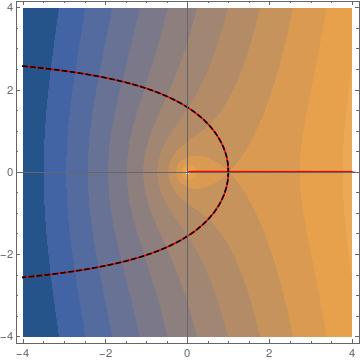}
\caption*{$\theta=0$}
\end{subfigure}
\begin{subfigure}{.22\textwidth}
\includegraphics[scale=.2]{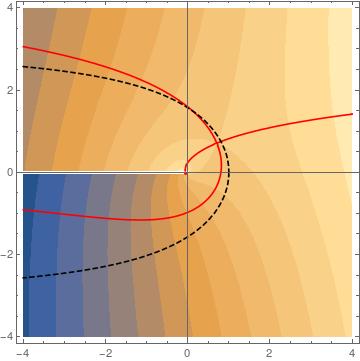}
\caption*{$\theta=\frac \pi 4$}
\end{subfigure}
\begin{subfigure}{.22\textwidth}
\includegraphics[scale=.2]{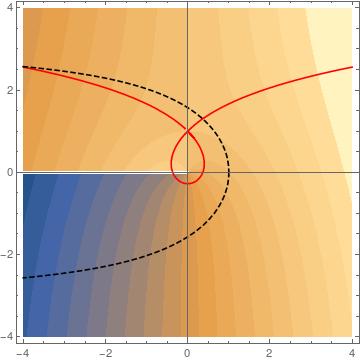}
\caption*{$\theta=\frac \pi 2$}
\end{subfigure}
\begin{subfigure}{.22\textwidth}
\includegraphics[scale=.2]{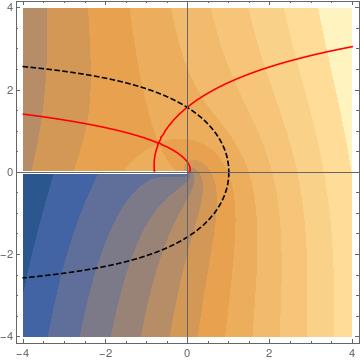}
\caption*{$\theta=\frac {3\pi}4$}
\end{subfigure}
\caption{{\it Steepest descent and ascent contours $\Im \varphi(x)=\Im \varphi(\e^{\i\theta})$ for the phase $\varphi(x)=x-\e^{\i\theta}\log x$ (red), and contour $C_2$ (black, dashed), for $\theta=i\frac{\pi}{4}$, $i=-3,...,3$. Level lines of $\Re\varphi$ are also shown. In all cases it is clear how to deform $C_2$ to the steepest descent contour in the vicinity of the saddle point, so that the contributions from the tails at infinity are exponentially smaller than the saddle point approximation.}}
\label{figuresaddle1}
\end{figure}

For all $|\theta|<\pi$, the contour $C_2$ can be deformed to the steepest descent contour $\Im \varphi(x)=\Im \varphi(\e^{\i\theta})$ in the vicinity of $x=\e^{\i\theta}$ in such a way that the main contribution to the integral for large $|\xi|$ comes from the neighborhood of the saddle point (see Fig. \ref{figuresaddle1}), and is computed by the gaussian integral;
\begin{align*}
\wh g(z-1)&\sim |\xi|\e^{-\frac\xi\epsilon\left(\log|\xi|-1+\i\theta\right)}\sum_{j\geq 0}\frac{(-1)^j}{j!(\epsilon|\xi|\e^{\i\theta})^j}\int_{\e^{\i\theta}+\e^{\i\frac{\pi+\theta}2}\R}\e^{\frac {|\xi|}\epsilon\frac{\e^{-\i\theta}}2 (x-\e^{\i\theta})^2}\d x
\\
&=\i\sqrt{2\pi\epsilon|\xi|}\e^{-\frac\xi\epsilon\left(\log\xi-1\right)}(1+\O(|\xi|^{-1})).
\end{align*}
Finally, we recall $\xi=\epsilon\left(z+\frac 12\right)$ and so $\sqrt{|\xi|}\e^{-\frac \xi\epsilon\left(\log\xi-1\right)}\sim\left(\frac{\epsilon z}\e\right)^{-z}$.

This completes the proof of the asymptotic for $g(z-1)$.

\paragraph*{Asymptotics for $\wh f$.}
Let us write $\xi:=\epsilon\left(z+\frac 32\right)$ and divide the contour $C_1$ in $C_{1}^{in}:=C_1\cap\{|x|\leq 1\}$ and $C_{1}^{out}:=C_1\cap\{|x|\geq 1\}$. Performing two different scalings $x\mapsto x|\xi|^{\pm 1}$ we have
\be
\wh f(z)=\frac{\e^{\frac\xi\epsilon\log|\xi|}}{|\xi|}\int_{|\xi|C_1^{in}}\e^{-\frac{|\xi|}{\epsilon}\left(\frac 1x+\e^{\i\theta}\log x\right)}\e^{\frac x{\epsilon|\xi|}}\d x+|\xi|\e^{-\frac \xi \epsilon \log|\xi|}\int_{|\xi|^{-1}C_1^{out}}\e^{\frac {|\xi|}\epsilon\left(x-\e^{\i\theta}\log x\right)}\e^{-\frac 1{\epsilon|\xi|x}}\d x
\ee
where $\xi=|\xi|\e^{\i\theta}$, $|\theta|<\pi$. Applying Fubini theorem, the first integral is
\be
\label{lastlast}
\int_{|\xi|C_1^{in}}\e^{-\frac{|\xi|}{\epsilon}\left(\frac 1x+\e^{\i\theta}\log x\right)}\e^{\frac x{\epsilon|\xi|}}\d x=\sum_{j\geq 0}\frac{1}{j!\epsilon^j|\xi|^j}\int_{|\xi|C_1^{in}}x^j\e^{-\frac{|\xi|}{\epsilon}\left(\frac 1x+\e^{\i\theta}\log x\right)}\d x
\ee
and the second one is also written similarly as in \eqref{last}.

We study each integral in the series in the right hand side of \eqref{lastlast} by the steepest descend method. The phase is $\varphi(x)=\frac 1x-\e^{\i\theta}\log x$, which has one saddle point at $x=\e^{-\i\theta}$. Expanding $\varphi(x)=\e^{\i\theta}(1-\i\theta)+\frac{\e^{3\i\theta}}2(x-\e^{-\i\theta})^2+\O((x-\e^{-\i\theta})^3)$ we see that the steepest descent direction is $-\frac{3\theta}2$. 

\begin{figure}[htbp]
\centering
\hspace{1cm}
\begin{subfigure}{.22\textwidth}
\includegraphics[scale=.2]{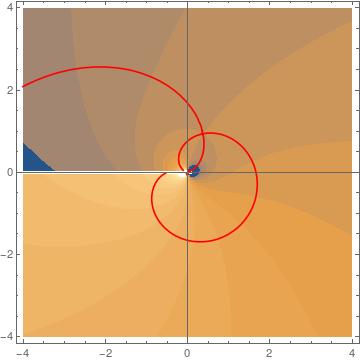}
\caption*{$\theta=-\frac {3\pi}8$}
\end{subfigure}
\begin{subfigure}{.22\textwidth}
\includegraphics[scale=.2]{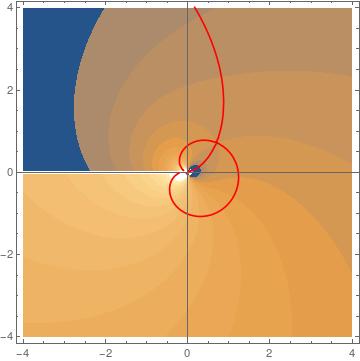}
\caption*{$\theta=-\frac \pi 4 \pi$}
\end{subfigure}
\begin{subfigure}{.22\textwidth}
\includegraphics[scale=.2]{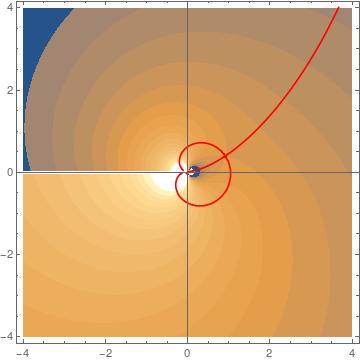}
\caption*{$\theta=-\frac \pi 8$}
\end{subfigure}
\begin{subfigure}{.22\textwidth}
\includegraphics[scale=.2]{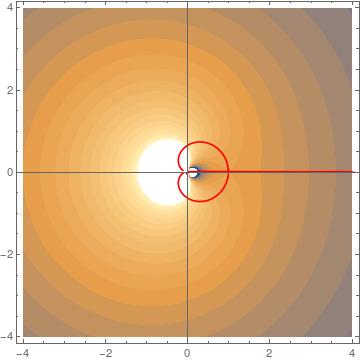}
\caption*{$\theta=0$}
\end{subfigure}
\begin{subfigure}{.22\textwidth}
\includegraphics[scale=.2]{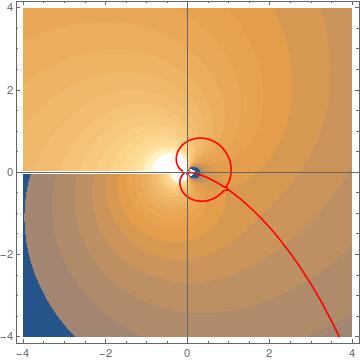}
\caption*{$\theta=\frac \pi 8$}
\end{subfigure}
\begin{subfigure}{.22\textwidth}
\includegraphics[scale=.2]{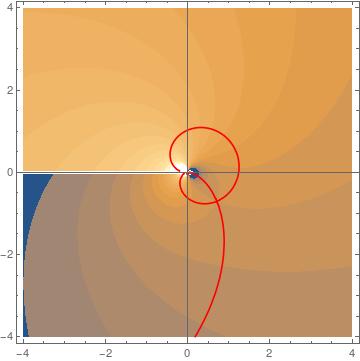}
\caption*{$\theta=\frac \pi 4$}
\end{subfigure}
\begin{subfigure}{.22\textwidth}
\includegraphics[scale=.2]{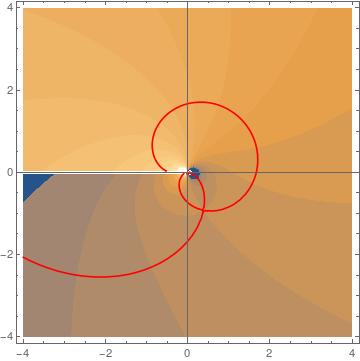}
\caption*{$\theta=\frac {3\pi}8$}
\end{subfigure}
\caption{{\it Steepest descent and ascent contours for the phase $\frac 1x+\e^{\i\theta}\log x$ (red) for $\theta=i\frac{\pi}{8}$, $i=-3,...,2$. Level lines of the real part of the phase are also shown. In all cases it is clear how to deform $|\xi|C_1^{in}$ to the steepest descent contour in the vicinity of the saddle point, so that the contributions from the tails at zero and infinity are exponentially smaller than the saddle point approximation.}}
\label{figuresaddle2}
\end{figure}

Let us restrict attention to 
\be
\label{rangetheta}|\theta|<\frac \pi 2.
\ee
The contour $C_1$ can be deformed so that $|\xi|C_1^{in}$ coincides with the steepest descent path in the vicinity of the saddle point $\e^{-\i\theta}$ (see Fig. \ref{figuresaddle2}), therefore giving the contribution
\be
\label{leading}
\frac{\e^{\frac\xi\epsilon\left(\log|\xi|-1+\i\theta\right)}}{|\xi|}\sum_{j\geq 0}\frac{(\e^{-\i\theta})^j}{j!\epsilon^j|\xi|^j}\int_{\e^{-\i\theta}+\e^{-\i\frac {3\theta}2}\R}\e^{-\frac{|\xi|}{\epsilon}\frac{\e^{3\i\theta}}2(x-\e^{-\i\theta})^2}\d x
=\frac{\sqrt{2\pi\epsilon}}{\xi^\frac 32}\e^{\frac\xi\epsilon\left(\log\xi-1\right)}(1+\O(|\xi|^{-1}))
\ee
where we recall that $\xi=\epsilon\left(z+\frac 32\right)$ so that $\xi^{-\frac 32}\e^{\frac\xi\epsilon\left(\log\xi-1\right)}\sim\left(\frac {\epsilon z}\e\right)^z$.
The contribution from the other term, relative to the contour $|\xi|^{-1}C_1^{out}$, is computed similarly as above for $g$ and is subleading with respect to \eqref{leading}, as long as we restrict to the range \eqref{rangetheta}.

This completes the proof of the asymptotics for $f$.

\section{Notions of geometry of moduli spaces}\label{appb}

The study of moduli spaces plays a prominent role in modern algebraic geometry, in particular for the relations with string theory and enumerative geometry, see e.g. \cite{Wi1993}.
In this appendix we review some main ideas in this topic to motivate all the technical terms employed in the introduction. The interested reader willing to further study these issues may consult the literature mentioned below.

\subsection{Moduli spaces of curves and Witten-Kontsevich theorem}

The moduli space $\mathcal M_{g,n}$ is the set of equivalence classes of pairs of a smooth (compact) Riemann surface $\mathscr C$  of genus $g$ together with a choice ${\bf P} = (P_1, \dots, P_n)$ of $n$ ``marked'' points. The equivalence between two such objects $(\mathscr C,\bf P)\sim (\widetilde{\mathscr C}, \widetilde {\bf P})$ is a  bi-holomorphic map that sends $P_i$ to $\wt P_i$.
The {\it stability condition} $2g+n>2$ (which is equivalent to the statement that the surface minus the points admits the unit disk as universal cover, i.e. it is {\it hyperbolic}) allows to show that this moduli space is parametrized by $3g-3+n$ complex parameters and has the structure of a non-compact complex {\it orbifold} (for the notion of orbifold see e.g. \cite[Chapter 13]{Thurston}, or \cite{Hain} for an exposition more related to moduli spaces of curves).

The compactification, $\Mgn$ was achieved via a construction by Deligne and Mumford \cite{DM}; this compactification allows for curves  with ``mild'' singularities (nodes, i.e. transversal self-intersections), provided that each connected component of a nodal curve  minus its nodes still satisfies the stability condition on its own.
The locus corresponding to smooth $\mathscr{C}$ is an open dense subset of $\Mgn$ identified naturally with $\mathcal M_{g,n}$ and in this sense we say that $\Mgn$ is a compactification of $\mathcal M_{g,n}$.

In $\Mgn$ there is an open-dense subset which has a manifold structure;  one can define the {\it cohomology}  of $\Mgn$ itself by  a suitable extension (which we don't discuss in this short review) of the ordinary notion on its  manifold  part.  

Of particular interest are the ``tautological (cohomology) classes''; these are the Chern classes, denoted by $\psi_i$  of certain  line bundles $\mathcal L_i$. These are the (complex) line bundles whose fiber at the point $(\mathscr C, {\bf P})\in \Mgn$ is  the co-tangent space of $\mathscr C$ at the point $P_i$ (a one-dimensional complex vector space). 
If $\Mgn$ were an ordinary (compact) complex manifold, then the integral of a volume form $\int_{\Mgn}\psi_1^{k_1}\cdots\psi_n^{k_n}$, $k_1+\dots+k_n = 3g-3+n$ (obtained from Chern classes of line bundles) would be an integer \cite{GriffithsHarris}. Since $\Mgn$ is instead an orbifold, the result is a rational number. These are the ``intersection numbers''  because of Poincar\'e\ duality \cite{GriffithsHarris}.

It came as a surprise to the mathematical community when Witten \cite{Wi1991} conjectured (on the basis of physical intuition about the equivalence of different approaches to 2D quantum gravity) the relation of these intersection numbers with the KdV hierarchy of integrable PDEs, expressed by saying that the exponential of \eqref{F} is a tau function of this hierarchy. The conjecture was proven by Kontsevich \cite{Ko1992} shortly after using the matrix model described in the introduction. The Witten-Kontsevich theorem is a crucial tool allowing to compute all intersection numbers using the KdV hierarchy equations as recursion relations for the intersection numbers (this was systematized in \cite{BeDuYa2016}).

\subsection{Moduli spaces of maps and Gromov-Witten theory}

An important generalization of the moduli spaces $\Mgn$ are the spaces $\Mgn(X,\beta)$; here $X$ is a smooth projective variety and $\beta\in H_2(X,\mathbb Z)$ representing the homology class of an immersion of an algebraic curve $\mathscr C \hookrightarrow X$.
Rather informally, these moduli spaces parametrize tuples $(\mathscr{C},P_1,\dots,P_n;f)$ where $\mathscr{C}$ is a Riemann surface of genus $g$ with simple nodal singularities at worst and with distinct marked points $P_i$ away from the nodes. The new ingredient is a holomorphic $f:\mathscr{C}\to X$ satisfying $f_*[\mathscr{C}]=\beta$.
Similarly as before, the objects are considered up to appropriate equivalence.

The rigorous construction of these moduli spaces is extremely complicated and technical, as it presents several new issues with respect to the moduli spaces of curves $\Mgn$. For the details we refer the reader to the introduction by Fulton and Pandharipande \cite{FP}.

The main motivation for considering such compact moduli spaces comes from \emph{enumerative geometry}. Indeed such spaces were introduced to address the computation of the number of inequivalent maps $\mathscr{C}\to X$ where the points $P_i\in \mathscr{C}$ are required to map to specified subvarieties $Y_i\subseteq X$. This generalizes classical old questions in algebraic geometry; for example, the problem of counting the number $N_1$ of lines in the plane passing through two points, or the number $N_2$ of conics in the plane through five points. In general, for $d\geq 1$, $N_d$  is the number of degree $d$ rational curves in the plane passing through $3d-1$ points. This classical problem was solved only recently by Kontsevich by providing a beautiful recursive formula for $N_d$ \cite{Kontsevich1994}, a cornerstone of modern enumerative geometry and Gromov-Witten theory.

Informally, the reasoning linking these moduli spaces to the enumerative geometry question above goes as follows and is mainly due to Kontsevich. Let $\gamma_i\in H^*(X,\mathbb Z)$ be the cohomology classes in $X$ which are Poincar\'e dual to the $Y_i$'s, and let $\ev_i:\Mgn(X,\beta)\to X$ be the ``evaluation maps'' sending $(\mathscr{C},P_1,\dots,P_n,f)$ to $f(P_i)$. Then the solution to this enumerative geometry question should be provided by an intersection number
\be
\label{GWdef}
\int_{\left[\Mgn(X,\beta)\right]}\ev_1^*\gamma_1\cdots\ev_n^*\gamma_n.
\ee
The expressions \eqref{GWdef} are called Gromov-Witten invariants. The classical example of degree $d$ rational plane curves through $3d-1$ points corresponds to $X=\P^2,\beta=(3d-1)[\P^2]$ and $\gamma_i$ Poincar\'e duals of a point in $\P^2$. 

There are several difficulties in making such a definition completely rigorous, the major of which is  a suitable construction of the homology class $\left[\Mgn(X,\beta)\right]$. As already remarked, the spaces $\Mgn(X,\beta)$ are generally singular and with smooth components of different dimensions, and thus the spaces lack a fundamental class in homology. A cornerstone in Gromov-Witten theory has been the construction due to Behrend and Fantechi \cite{Behrend1996} of a \emph{virtual} fundamental class $[\Mgn(X,\beta)]$ of the appropriate dimension; then the intersection number \eqref{GWdef} (denoting the pairing of the cohomology class $\prod_{i=1}^n\ev_i^*\gamma_i$ with the virtual fundamental class by an integration, as if the spaces were smooth) is well defined.

The Gromov-Witten invariants appearing in \eqref{FP1} also involve integration of psi-classes; these are obtained in general from the classes $\psi_i$ on $\Mgn$ by pull-back via the forgetful map $\Mgn(X,\beta)\to\Mgn$ (defined essentially by disregarding the map $f$).
Insertion of psi-classes is in some sense related to the enumerative problem of counting maps $\mathscr{C}\to X$ with incidence as well as \emph{tangency} conditions, see e.g. \cite{Graber}.

Finally we specialize to the case considered in this paper, see \eqref{FP1}, in which $X=\P^1$ is the the complex projective line, i.e. the Riemann sphere. The cohomology $H^*(\P^1,\mathbb Z)$ is $\mathbb Z[1,\omega]/\left\langle\omega^2=0\right\rangle$ where $\omega\in H^2(\P^1,\mathbb Z)$ is normalized the volume (K\"{a}hler) form, $\int_{\P^1}\omega=1$. By the informal discussion above, when all $\gamma_i=\omega$,  the images of $P_i$ are fixed ($\omega$ is the Poincar\'e dual to the class of a point), whence the adjective ``stationary'' for such case. Since $H_2(\P^1,\mathbb Z)\simeq \mathbb Z$ is generated by the fundamental class $[\P^1]$, for the degree we use the simpler notation $d\in\mathbb Z$, implying $\beta=d[\P^1]$; we also note that $\Mgn(\P^1,d)$ is empty unless $d\geq 0$.

\end{document}